\documentclass[10pt, journal,comsoc]{IEEEtran}

\usepackage[T1]{fontenc}% optional T1 font encoding

\ifCLASSINFOpdf
\else
\fi

\usepackage{amsmath}

\usepackage{xcolor,colortbl}
\usepackage{float}
\usepackage{graphicx}
\usepackage{subfigure}
\usepackage{multirow}
\usepackage{tcolorbox}
\usepackage{xcolor}
\usepackage{colortbl}
\usepackage{setspace}
\usepackage{color}
\usepackage{algpseudocode}
\usepackage{hyperref}
\usepackage{framed}
\usepackage{amsmath}
\usepackage{pifont}
\usepackage{enumitem}
\usepackage{balance}
\usepackage{verbatim}
\usepackage{url}
\usepackage[T1]{fontenc}
\usepackage[utf8]{inputenc}
\usepackage[font=small,labelfont=bf,tableposition=top]{caption}
\usepackage{booktabs}
\usepackage{threeparttable}
\usepackage[ruled,vlined,linesnumbered]{algorithm2e}

\definecolor{grey}{rgb}{0.9,0.9,0.9}
\definecolor{lightgreen}{HTML}{bae4b3}
\definecolor{lightgrey}{HTML}{f0f0f0}
\definecolor{mygreen}{HTML}{31a354}
\definecolor{mygray}{HTML}{666666}

\hypersetup{
	colorlinks=true,      % false: boxed links; true: colored links
	linkcolor=black,       % color of internal links
	citecolor=black,    % color of links to bibliography
	filecolor=cyan,       % color of file links
	urlcolor=black          % color of external links
}

\newcommand{\wu}[1]{\mytodocyan{[wu: #1]}}

\newcommand{\mytodocyan}[1]{\textcolor{cyan}{\ding{46}~{\sf}~#1}}

\newcommand*{\eg}{e.g., }
\newcommand*{\mycode}{\fontfamily{lmtt}\selectfont}

\newcommand*{\ie}{i.e., }

\usepackage[cmintegrals]{newtxmath}
\begin{document}
\title{Will Dependency Conflicts Affect My Program's Semantics?}

\author{Ying Wang, Rongxin Wu, Chao Wang, Ming Wen, Yepang Liu,  Shing-Chi Cheung~\IEEEmembership{Senior Member,~IEEE}, \\Hai Yu, Chang Xu~\IEEEmembership{Senior Member,~IEEE}, Zhiliang Zhu~\IEEEmembership{Member,~IEEE}

% <-this % stops a space
\thanks{Ying Wang, Chao Wang, Hai Yu, and Zhiliang Zhu are with the Software College, Northeasthern University, China. E-mail: wangying@swc.neu.edu.cn, wangc\_neu@163.com, \{yuhai, zzl\}@mail.neu.edu.cn.}% <-this % stops a space
\thanks{Rongxin Wu is with Department of Cyber Space Security, Xiamen University, China. E-mail: wurongxin@xmu.edu.cn.} \thanks{Ming Wen is with School of Cyber Science and Engineering, Huazhong University of Science and Technology, China. E-mail: mwenaa@hust.edu.cn.}
\thanks{Yepang Liu is with Department of Computer Science and Engineering, Southern University of Science and Technology, China. E-mail: liuyp1@sustech.edu.cn.}
\thanks{Shing-Chi Cheung is with Department of Computer Science and Engineering, The Hong Kong University of Science and Technology, China. E-mail: scc@cse.ust.hk.}
\thanks{Chang Xu is with State Key Lab for Novel Software Technology and Department of Computer Science and Technology, Nanjing University, China. E-mail: changxu@nju.edu.cn.}
\thanks{Manuscript received May 9, 2020; revised August 26, 2020.}}


\markboth{IEEE Transactions on Software Engineering,~Vol.~6, No.~6, May~2020}%
{Wang \MakeLowercase{\textit{et al.}}: Will Dependency Conflicts Affect My Program's Semantics?}



\maketitle

\begin{abstract}
Java projects are often built on top of various third-party libraries.
If multiple versions of a library exist on the classpath, 
JVM will only load one version and shadow the others, 
which we refer to as \textit{dependency conflicts}.    
This would give rise to \textit{semantic conflict} (SC) issues, 
if the library APIs referenced by a project 
have identical method signatures but inconsistent semantics 
across the loaded and shadowed versions of libraries. 
SC issues are difficult for developers to diagnose in practice,
since understanding them typically requires domain knowledge.
Although adapting the existing test generation technique for dependency conflict issues, \textsc{Riddle}, to detect SC issues is feasible,
its effectiveness is greatly compromised. 
This is mainly because 
\textsc{Riddle} randomly generates test inputs, 
while the SC issues typically require specific arguments in the tests to be exposed. 
To address that, we conducted an empirical study of 75 real SC issues to understand the characteristics of such specific arguments in the test cases that can capture the SC issues.
Inspired by our empirical findings, we propose an automated testing technique \textsc{Sensor}, which synthesizes test cases using ingredients from the project under test to trigger inconsistent behaviors of the APIs with the same signatures in conflicting library versions.
Our evaluation results show that \textsc{Sensor} is effective and useful: it achieved a $Precision$ of 0.803 and a $Recall$ of 0.760 on open-source projects and a $Precision$ of 0.821 on industrial projects; it detected 150 semantic conflict issues in 29 projects, 81.8\% of which had been confirmed as real bugs.
\end{abstract}


\IEEEpeerreviewmaketitle

\vspace{-3mm}
\section{Introduction}
\label{sec:Introduction}
%Reusing code has long been seen as an important approach to reduce the cost and increase the quality of software system~\cite{jezek2015java}.
%One of the reuse success stories is the use of third-party libraries in Java projects. 
%This is facilitated by a combination of social factors such as the existence of vibrant open source communities and commercial product ecosystems~\cite{thung2013automated}.
%While the benefits of reusing code, an often overlooked risk are dependency conflicts due to Java language features like name spaces, class loading mechanism, etc~\cite{artho2012software}.
%If multiple versions of the same library or class are present on a \emph{classpath}, only one version will be loaded while the others will be shadowed~\cite{liang1998dynamic}. 
%As API changes in libraries evolve independently, a DC (\underline{D}ependency \underline{C}onflict) issue occurs if these versions coexisting in one project are incompatible.
%In such scenario, unexpected program behaviors will happen when a project references the incompatible methods of shadowed versions at runtime.

\IEEEPARstart{B}{uilding} software projects on top of third-party libraries is a common practice to save development cost and improve software quality \cite{ bao2018inference, jezek2015java, teyton2014study, macho2018automatically}.
However, the heavy dependencies on third-party libraries often induce dependency conflict issues \cite{wang2018dependency}. 
When multiple versions of the same library class are present on the {\mycode classpath}, the Java class loader will load only one version and shadow the others~\cite{liang1998dynamic}. 
If the loaded version has inconsistent implementations with the intended but shadowed versions, dependency conflict issues will occur, inducing risks of runtime exceptions or unexpected program behaviors.%\yepang{Can we be more specific and explain the meaning of ``incompatible'' here?}

\begin{figure}[t!]
	%\vspace{-3mm}
	\centering
	\includegraphics[width=0.40\textwidth]{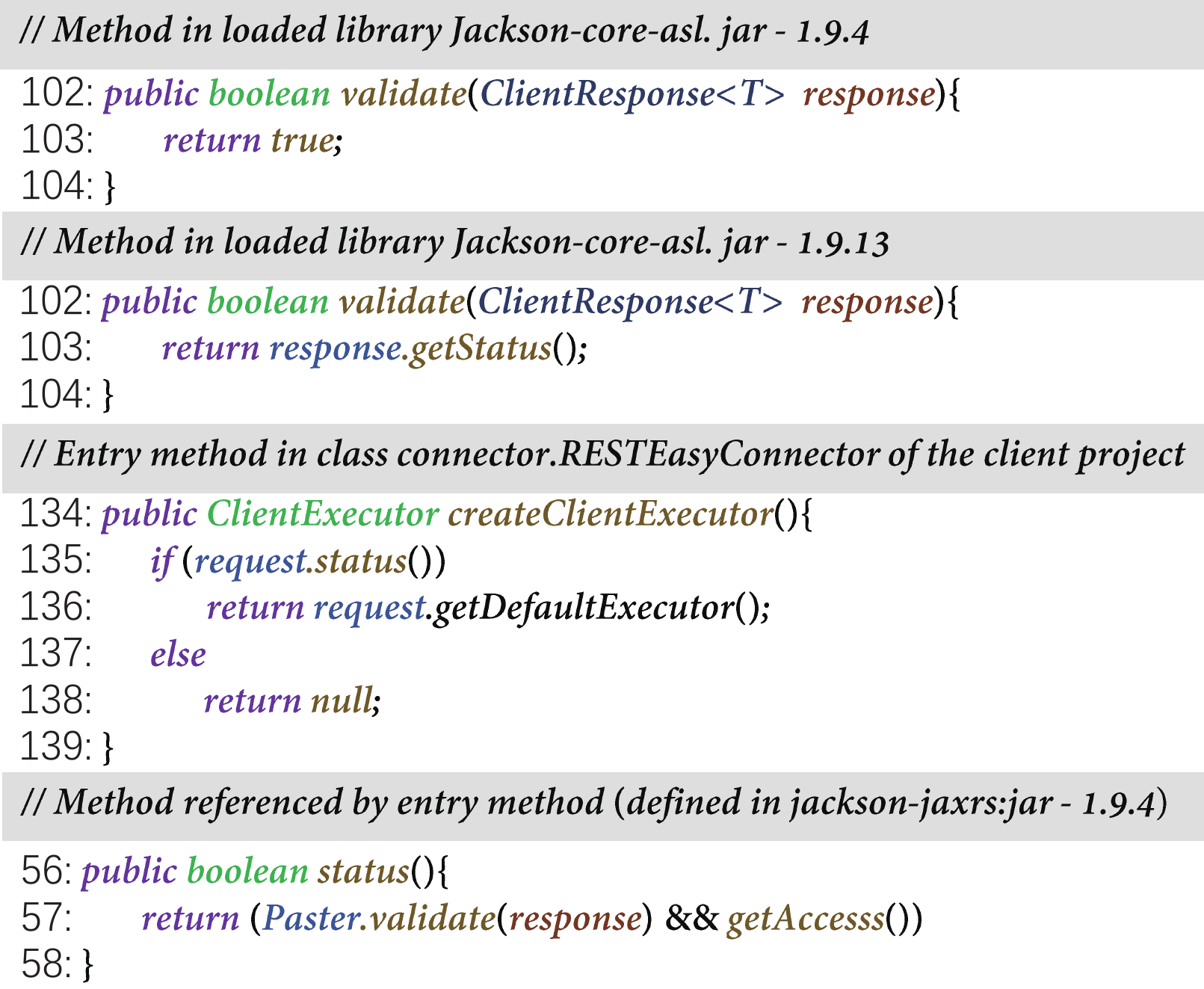}
	%	\vspace{-4mm}
	\caption{{\mycode Issue \#214}~\cite{Issue214} in the project {\mycode Openstack-java-sdk 3.2.5}}	
	\label{fig_11}
	\vspace{-8mm}
\end{figure}

The state-of-the-art techniques \cite{wang2018dependency, WANG2019STACK} for detecting dependency conflict issues mainly focus on specific categories of the issues, such as {\mycode ClassNotFoundException} and {\mycode NoSuchMethodError}, which happen when the loaded library versions do not cover all the APIs referenced by the client projects. 
One limitation of these techniques is that they cannot identify the dependency conflict issues that arise from referencing those APIs with identical method signatures but inconsistent behaviors across multiple library versions \cite{wang2018adversarial, WANG2019STACK}. 
%this will lead to and cannot be detected by the existing DC issue detection techniques \cite{wang2018adversarial, WANG2019STACK}.
We refer to such issues as \underline{S}emantic \underline{C}onflict issues (SC issues for short).
Figure \ref{fig_11} gives a real example of SC issues.
On the {\mycode classpath} of the project {\mycode Openstack-java-sdk 3.2.5}, there are two versions of the library {\mycode Jackson-core-asl}, namely, \textit{Version 1.9.4} and \textit{Version 1.9.13}.
In the example, Java class loader loads \textit{Version 1.9.13} but shadows \textit{Version 1.9.4}.  
As shown in the code snippet, 
the method {\mycode createClientExecutor()} in the project will transitively invoke {\mycode validate(ClientResponse)} of the library {\mycode Jackson-core-asl}.
However, the implementations of {\mycode validate(ClientResponse)} are semantically inconsistent between the two
versions. 
The project was originally designed to use \textit{Version 1.9.4} of {\mycode Jackson-core-asl}, which is unfortunately shadowed.  
Although there will be no runtime exceptions in such cases, the semantic inconsistency of library method implementations will inappropriately affect the variable states of the client project via the invocation of the concerned methods, 
leading to unexpected program behaviors.

%Many code changes are very subtle and it cannot be expected that all developers understand the effects of seemingly minor API changes~\cite{nguyen2014statistical}, 
%let alone that they may not be aware of the replacement of library version caused by DC issues.
%As such behavioral differences would not break API invocations and are difficult for developers to trigger them using tests with suitable inputs, they bring great hidden trouble to the maintenance of the project.

%Taking a real {\mycode Issue \#214}~\cite{Issue214} for example, multiple versions of library {\mycode Jackson}{\mycode -core-asl} exist in project {\mycode Openstack-java-sdk-3.2.5}.
%Based on build tool's dependency management strategy, only {\mycode Jackson-core-asl:1.9.4} can be loaded, and {\mycode Jackson-core-asl:1.9.13} will be shadowed.
%From its dependency tree, developers found that their project expects to reference {\mycode Jackson-core-asl:1.9.13} on one dependency path. Due to such a DC issue, {\mycode Openstack-java-sdk-3.2.5} is forced to invoke the methods included in the actual loaded version {\mycode Jackson-core-asl:1.9.4}.
%As shown in Figure~\ref{fig_11}, an expected callee {\mycode Utf8StreamParser.validate()} defined in shadowed library has different implementations from actual referenced callee with the same signature included in loaded library.
%Although runtime exceptions will not be thrown, such differences can change the variable states of the client project, which leads to unexpected semantic program behaviors.

%\vspace{-3mm}

%Moreover, SC issue is very difficult for developers to diagnose. 

%Semantic behavior changes often occur with dependency conflicts.
%It affects program's functionality and is difficult for developers to diagnose.
%In a {\mycode PR}~\cite{PRIssue214} of the above {\mycode Issue \#214}, a developer complained that

SC issues arise from code changes of API implementations, which are common in popular libraries. 
Many of these changes are too subtle for developers to understand their effects on program semantics \cite{jezek2015java}. Existing test suites may not help effectively expose such differences neither. 
As such, SC issues are difficult to diagnose. 
For example, a developer left the following comment in the pull request~\cite{PRIssue214} of the aforementioned issue:

%\vspace{-4mm}
{\leftskip0.3cm\relax
	\rightskip0.4cm\relax
	\emph{``I have encountered these types of semantic inconsistency issues lots of times when dealing with dependency conflicts.
		%As far as I understand Java, there is no version checking at runtime.
		When such issues happen, signature changes can be detected by static analyzers.
		However, semantic changes would be more difficult to detect. Empirically, developers diagnose them by reading the git history of the library or dynamic testing.''}
	\par}

%Reporting APIs with any change of code structure would induce a large number of candidate SC issues.
Detecting SC issues typically requires rich domain knowledge to discern the subtle differences in API implementations, which is a non-trivial task.
Therefore, an automated technique to detect SC issues is highly desirable. 
We note that the most relevant and recent technique is \textsc{Riddle} \cite{WANG2019STACK}. 
It was designed to verify dependency conflict issues caused by the missing of classes or methods. 
\textsc{Riddle} can generate tests to drive the execution of a client project towards the target call sites that could induce dependency conflict issues. 
While it seems possible to adapt \textsc{Riddle} using the idea of differential testing (i.e., comparing the runtime behavior of a target API across multiple library versions) to detect SC issues, our trial on 70 open-source projects shows that this technique is not that effective as anticipated (see Section~\ref{sec:Motivation Example}). 
%We \textsc{Riddle} on 70 open source projects indicates that 
%the effectiveness of such adaptation is greatly compromised (see Section~\ref{sec:Motivation Example}). 
It is mainly because merely reaching the call sites of a target API and invoking it with random arguments can hardly trigger the inconsistent behaviors of the API across different versions.
As such, many SC issues, whose manifestation requires specific arguments (referred to as \textit{divergence arguments} in this paper), cannot be effectively exposed. %\scc{Consider to find a better term for valid arguments.} 
This motivates us to design a more effective testing technique to detect SC issues.

%This motivates us to develop an automated testing technique to detect SC issues.
%triggering SC issue requires certain specific inputs, 
%Unfortunately, \textsc{Riddle} generates test inputs randomly. %is a random test generation technique that can only generate random inputs.   
%This motivates us to conduct an empirical study to understand 
%the characteristics of the test inputs that can expose inconsistent semantic behaviors.  

As discussed above, an obvious challenge in detecting SC issues via testing is to generate divergence arguments to trigger inconsistent API behaviors across different library versions. To address this challenge, %we conducted an empirical study by collecting a benchmark of test cases that successfully capture 75 SC issues, to understand the characteristics of the valid arguments in the test cases that can expose inconsistent semantic behaviors.
we performed an in-depth study of 75 real SC issues collected from open-source Java projects to understand the characteristics of divergence arguments in the test cases that could expose these issues.
%Our investigation based on a benchmark of test cases that successfully capture 75 SC issues, reveals several interesting findings. 
The study revealed several interesting findings.
First, to generate class instances as test inputs for detecting SC issues, %\civi{I cannot understand the previous sentence, what are ``valid class instances as test inputs''?}
almost all (98.5\%) the object constructors take at least one argument,
and, for 97.8\% of these constructors, 
at least one of their arguments has specific values that can hardly be generated by random techniques such as \textsc{Riddle}. 
Second, we observed three common patterns to produce divergence arguments for object constructors in the test cases. 
Third, we found that, for 56.9\% of our analyzed object constructors, 
their divergence arguments can be directly obtained from the source code of the client project. 
Fourth, for the constructors in the test cases whose arguments cannot be found in the source code of the client project, we replaced the arguments with other compatible values that can be found in the client project's source code and discovered that 37 out of 58 (63.8\%) such revised test cases could still capture SC issues.
%The above results revealed that 
%\civi{what does ``valid arguments'' mean here? valid in terms of what?}
%We found that 37 out of the 58 revised test cases (63.8\%) could still capture SC issues.%\yepang{This sentence is hard to follow.}
%\wu{please see whether the findings are correct and complete the rest one.}

%Inspired by our empirical findings, we designed an automated testing technique, \textsc{Sensor}, to detect SC issues for Java projects.
%As the empirical results reveal that object constructors in the test cases need specific arguments to trigger SC issues, and most of which can be found in the source code of client project,
%we propose to synthesize object constructors of the arguments required to invoke the APIs under test. 
%Specially, we study the set of legitimate API usages and values of their arguments 
%in the client project's source code for the synthesis of a constructor.   
%We refer to the set as the constructor's \emph{invocation context}. 

Inspired by our empirical findings, 
our idea of generating divergence arguments for triggering SC issues 
is to synthesize these arguments 
from the source code of client projects.
Specifically, we synthesize an object constructor of a divergence argument 
by distilling the set of legitimate API usages and the values of its arguments from the source code. 
We refer to the set as the constructor's \emph{invocation context}.
We implemented our idea into an automated testing technique, \textsc{Sensor}. 
Given a client project to analyze, \textsc{Sensor} first extracts the invocation contexts of each object constructor from the source code and leverages them to construct a pool of class instances.
%Given a project to analyze, \textsc{Sensor} first extracts the invocation contexts of object constructors for those library classes that have undergone implementation changes from the source code of the client project.
Combining a seeding strategy of class instances with \textsc{Evosuite}, it then generates tests to trigger the concerned library APIs and checks whether they behave consistently across different versions. 
In our approach, \textsc{Sensor} does not simply report all detected behavioral inconsistencies as bugs.
Instead, it pinpoints the differences in variable states of a project under analysis and provides such fine-grained information to help developers further diagnose SC issues.
We evaluated \textsc{Sensor} using 92 open-source projects on {\mycode GitHub} and 10 industrial projects from  {\mycode Neusoft} Co. Ltd (SSE: 600718)~\cite{Neusoft}. \textsc{Sensor} achieved a $Precision$ of 0.803 and a $Recall$ of 0.760 on open-source projects, and a precision of 0.821 on industrial projects.
\textsc{Sensor} detected 150 real SC issues from 29 open-source projects. %\yepang{Did Sensor find any real SC issues in the 10 industrial projects?}
We reported these issues to the developers of the corresponding projects and detailed the issues' impact on program behaviors.
So far, 81.8\% of our reported issues have been confirmed by the developers as real bugs, and 85.2\% of the confirmed issues have been fixed quickly.
%We consolidated the 146 issues and submitted 32 issue reports to the developers of the corresponding projects. In the issue reported, we detailed the issues' impacts on the program semantics of these projects.
%So far, 114 reported issues reported in twenty-five reported issues have been confirmed as real SC issues.
%Developers have quickly fixed 19 of them.
Most of the confirmed issues are from popular projects such as {\mycode Rest-assured}~\cite{Rest-assured} and {\mycode Java-design-patterns}~\cite{Java-design-patterns}. %\yepang{Give links to the two repos?}
From the feedback on our reported issues (see Section~\ref{sec:Usefulness}), we observed that developers acknowledged the pervasiveness of SC issues and the necessity of a testing technique to diagnose such issues.
They also expressed great interests in using \textsc{Sensor}.
These results demonstrate the effectiveness and usefulness of \textsc{Sensor}.
In summary, we make four major contributions in this paper:

\begin{itemize}[leftmargin=*]
	\item An empirical study of 75 real SC issues for exploring the characteristics of test cases that can expose SC issues.
	\item A fully automated technique, \textsc{Sensor}, for detecting SC issues.% for Java projects.
	\item A benchmark dataset for assessing \textsc{Sensor} and similar approaches
	for detecting the issues induced by semantic inconsistencies of
	library APIs across different versions.
	\item A systematic analysis and discussions of SC issues' impacts on program behaviors. 
\end{itemize}

Our tool and dataset are available at: \underline{https://sensordc.github.io/}. 
\vspace{-1mm}
\section{Preliminaries}
\label{sec:PRELIMINARIES}

%\vspace{-5mm}
\begin{figure*}[ht]
	\centering
	\includegraphics[width=0.85\textwidth]{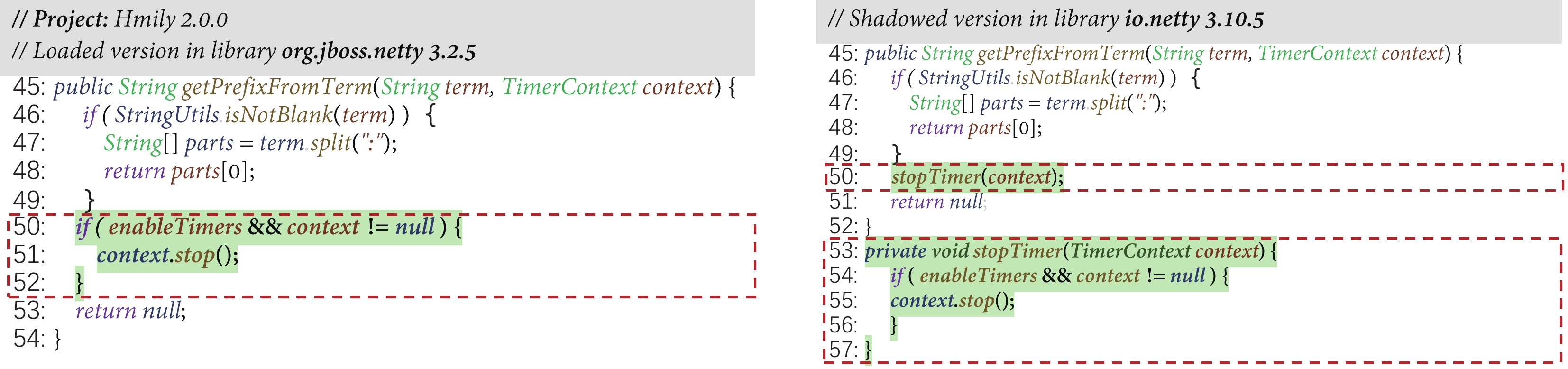}
	%\vspace{-4mm}
	\caption{A motivating example}
	\label{fig_R}
	\vspace{-5mm}
\end{figure*}

\subsection{Motivation}

%If a client project contains a conflicting API pair,
%it does not necessarily introduce a SC issue, 
%since the semantics of this API pair may still be consistent.
To roughly estimate the scale of SC issues, we statically detected the semantic inconsistency of the conflicting API pairs %with respect to their \textit{code structures} 
by comparing their \emph{code structures} in terms of call graphs and control flow graphs.  
%is to analyze their implementation differences in terms of their call graphs and control flow graphs, 
%which are collectively referred to as \textit{code structures}. 
%Following this approach, we investigated the pervasiveness of the potential SC issues.
We first collected 1,654 Java projects from {\mycode GitHub} 
based on two criteria:
(1) it has achieved over 50 stars or forks (popularity); and
(2) it is built on the {\mycode Maven} platform.
Then, we compared the code structures of the conflicting API pairs in these projects and labeled them as potential SC issues if their code structures are different.
%to identify potential SC issues. 
The results showed that  
73.1\% of the projects contain at least one potential SC issue.
Each of them contains on average 20 conflicting library API pairs that potentially cause
SC issues.

%However, difference with respect to code structures does not necessarily introduce a SC issue.
The static analysis of SC issues based on different code structures can be highly imprecise for some projects. 
%\scc{What are code structures? control flow structures? dataflow structures?}
%it is still difficult to precisely determine true positives.
Figure~\ref{fig_R} illustrates a false positive SC issue found by the static approach.
There are two versions of the class {\mycode Netty.bootstrap}{\mycode .ServerBootstrap} 
on the {\mycode classpath} of the project {\mycode Hmily-2.0.0}~\cite{Hmily}, 
which are included by the libraries  {\mycode org.jboss.netty 3.2.5}  and {\mycode io.netty 3.10.5}, respectively. 
Due to {\mycode Maven}'s \emph{first declaration wins} strategy, only the 
method {\mycode getPrefixFromTerm()} declared in the library {\mycode org.jboss.netty 3.2.5}  
is loaded and invoked by the client project.
Although the call graphs of these two versions differ, the method in {\mycode io.netty 3.10.5} was simply a code refactoring in {\mycode org.jboss.netty 3.2.5} that did not affect the program semantics. %\scc{It is a bit controversial to argue that they have different control flow graphs. Do we mean syntactically different?}
%\civi{are there any other cases instead of refactoring?}

Validation of SC issues is non-trivial. It requires domain knowledge to understand the implementations of the client project and its libraries.
%deep understanding of their client project's implementation, dependency libraries and domain knowledge. 
%Such level of understanding is unattainable even by some developers as shown in the aforementioned example bug report ~\cite{PRIssue214}. %\ying{Why was this reference broken?}
This motivates us to validate SC issues using automatically generated tests. 

\vspace{-3mm}
\subsection{Problem formulation}
\label{sec:Problem_Formulation}

%\wu{Can we change this section into ``preliminary'' section, which only introduce those necessary terminology. Actually, we do not need to specifically formulate the problem. The problem is clear via the previous motivation example. }

%Given a project with a dependency conflict $I_i$ due to its dependency on multiple versions of the class $C_i$ or library $Lib_i$.

To formulate our research problem, we introduce the following concepts.
In particular, we let $C_{i}'$ be a shadowed class version and $C_{i}$ be the actually-loaded class version, and use $\underline{C::m}$ to denote an API $m$ of class $C$.

%\textbf{Definition 1. (Semantic conflict):} 
%If unexpected semantic behaviors arise, when an API defined in the shadowed class version $C_i'$ 
%and referenced by a host project is forced to be replaced by another API in the actually-loaded class version $C_i$, 
%then we define it as a semantic conflict. 
%\scc{Please rewrite. Unparsable!}

\textbf{Definition 1. (Conflicting API pair):} Let $C_i'\!::\!m_k$ be an API included in the shadowed class version  and referenced by the client project $\mathcal{H}$, and $C_i\!::\!m_k$ be an API belonging to the actually-loaded class version,
where $m_k$ represents the method signature.
%If $C_i'\!::\!m_k$ can be replaced by $C_i\!::\!m_k$ without breaking the invocation, 
If $C_i'\!::\!m_k$ and $C_i\!::\!m_k$ share the same signature, 
we consider $C_i'\!::\!m_k$ and $C_i\!::\!m_k$ as a pair of conflicting APIs, which is denoted as $RP\langle C_i^\prime\!::\!m_k, C_i\!::\!m_k\rangle$.
Conflicting class versions $C_i$ and $C_i'$ caused by a dependency conflict issue, may introduce a set of conflicting API pairs. 
We denote the set of conflicting API pairs as $R\langle C_i, C_i'\rangle$.
Specially, if there are implementation differences between $RP\langle C_i^\prime\!::\!m_k, C_i\!::\!m_k\rangle$, we consider it as an \textit{isomerous conflicting API pair}.

%\textbf{Definition 1. (Conflicting API pair):} Let $C_i^\prime\!::\!m_k$ be an API included in the shadowed class version  and referenced by the client project $\mathcal{H}$, and $C_i^\prime\!::\!m_k$ be an API belonging to the actually-loaded class version,
%where $m_k$ represents the a method signature.
%If $C_i^\prime\!::\!m_k$ can be replaced by $C_i^\prime\!::\!m_k$ without breaking the invocation, we consider $C_i^\prime\!\!::\!\!m_k$ and $C_i\!::\!m_k$  as a pair of conflicting APIs, which is denoted as $RP\langle C_i^\prime\!::\!m_k, C_i\!::\!m_k\rangle$.
%For the dependency conflict $I_i$, conflicting class versions $C_i$ and $C_i'$ may introduce a set of conflicting API pairs. We denote the conflicting API pair set as $R\langle C_i, C_i'\rangle$.

\textbf{Definition 2. (Original dependency path):} 
For each API $C_i'\!::\!m_k$ included in a shadowed class version and referenced by a class $C_1$ in the client project, we define any path $Ep = \langle C_1\!::\!m_1, \cdots, C_{i-1}\!::\!m_{k-1}, C_{i}'\!::\!m_{k} \rangle$ as its original dependency path, where $C_1\!::\!m_1$ represents an entry method in the class $C_1$ of the client project indirectly referencing the method $C_{i}'\!::\!m_{k}$ along $Ep$.

\textbf{Definition 3. (Actual dependency path):} Suppose that $RP\langle C_i^\prime\!::\!m_k, C_i\!::\!m_k\rangle$ is a conflicting API pair. For each original dependency path $Ep$ with respect to API $C_{i}'\!::\!m_{k}$, we define $Fp = \langle C_1\!::\!m_1, \cdots, C_{i-1}\!::\!m_{k-1}, C_{i}\!::\!m_{k} \rangle$ as the corresponding actual dependency path, as the build environment enforces the interactions between entry method $C_1\!::\!m_1$ in the class $C_1$ of the client project and API $C_{i}\!::\!m_{k}$ included in the actually-loaded class $C_i$ along $Fp$.
Note that, $Ep$ and $Fp$ share the subpath from entry method $C_1\!::\!m_1$ to $C_{i-1}\!::\!m_{k-1}$. 
%Note that path from entry method $I(m_t, \mathcal{H})$ to $I(m_k, C_t)$ in $Ep_{tk}$ and $Fp_{tk}$ must be the same.

\textbf{Problem:}
Given a project with a set of conflicting API pairs $R\langle C_i, C_i'\rangle$, our research problem is %to detect SC issues using automated test generation technique. %to capture the semantic inconsistencies of a client project?
%Specifically, we should generate tests 
how to design an automated test generation technique to trigger the executions of each isomerous conflicting API pair $RP\langle C_i^\prime\!::\!m_k, C_i\!::\!m_k\rangle \in R\langle C_i, C_i'\rangle$ along their original and actual dependency paths, respectively, thereby identifying their impacts on the client project's program behaviors.

\vspace{-4mm}
\subsection{Challenges}
\label{sec:Motivation Example}

\textsc{Riddle} is the state-of-the-art technique that generates tests to detect dependency conflict issues in projects where the loaded library versions fail to 
cover all the referenced APIs based on their method signatures \cite{WANG2019STACK}. 
However, this technique is not applicable to detecting SC issues, 
since SC issues arise from referencing the APIs with identical method signatures 
but inconsistent behaviors across multiple library versions. 

%
%\begin{figure}[ht]
%	\vspace{-4mm}
%	\centering
%	\includegraphics[width=0.505\textwidth]{./figure/F_3.pdf}
%	\vspace{-9mm}
%	\caption{An example code for illustrating challenge}	
%	\label{fig_w}
%	\vspace{-3mm}
%\end{figure}

\textsc{Riddle} generates tests forcing the program execution along the path that invokes the shadowed library APIs. We may adapt the mechanism to detect SC issues. 
For example, after identifying the conflicting library APIs for SC issues, 
we can use \textsc{Riddle} to generate tests 
to drive the program to execute along the path
from an entry method to the conflicting library APIs.
Then, we can execute the generated test 
using the shadowed version and the loaded version, respectively, 
and compare their test outcomes to check the semantic inconsistency. %\scc{Multiple forms of the same concept? inconsistent semantic behavior, inconsistent behavior, inconsistent semantics, ... possibilities of unifying these terms?}
However, based on our trials on 70 Java projects with \textsc{Riddle}, 
we observed that this approach cannot effectively detect SC issues.
%This is because,
%the majority of SC issues require specific test inputs to trigger (See Section \ref{sec:Empirical}),
%while \textsc{Riddle} is essentially a random testing approach 
%and less effective in generating such inputs. 
%
%since \textsc{Riddle} is less effective in producing valid class instances 
%which are essential to capture the semantic behavior. 

\begin{figure}[t!]
	\vspace{-2mm}
	\centering
	\includegraphics[width=0.42\textwidth]{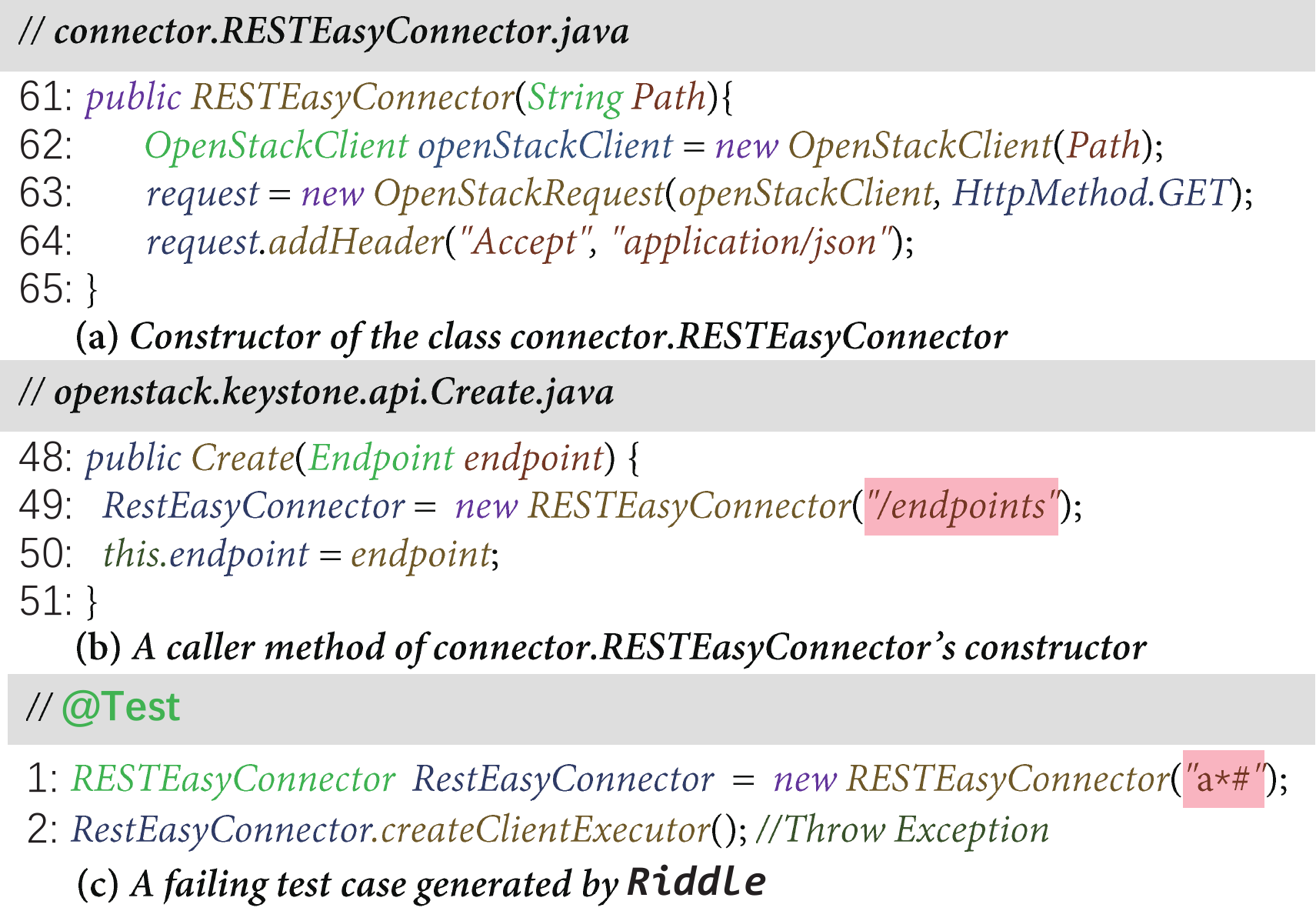}
	%\vspace{-3mm}
	\caption{Tests generated by {\mycode Riddle} for the example issue \#214~\cite{Issue214}}	
	\label{fig_2}
	\vspace{-4mm}
\end{figure}

%Figure \ref{fig_w} and \ref{fig_2} show an example code snippet of SC issue 
%and the test case generated by \textsc{Riddle} respectively. 
%The client project directly depends on {\mycode a-1.0.Jar} and {\mycode b-2.0.Jar}, 
%and transitively depends on {\mycode b-1.0.Jar}. 
%Since the libraries {\mycode b-2.0.Jar} and {\mycode b-1.0.Jar} 
%are the same library in different versions,
%this will lead to a dependency conflict. 
%Due to {\mycode Maven}'s \emph{nearest wins strategy}, 
%{\mycode b-2.0.Jar} shadows  {\mycode b-1.0.Jar}. 
%The library API  {\mycode SevConfig} {\mycode .getSize()} referenced by the client project
%has different implementations in version {\mycode b-2.0.Jar} and {\mycode b-1.0.Jar}, 
%and the intended library version is {\mycode b-1.0.Jar} which is shadowed.
%Thus, this will lead to a potential SC issue.  
Consider the SC~issue \#214~\cite{Issue214} described in Section~\ref{sec:Introduction}.
To trigger the conflicting library API {\mycode validate(ClientResponse)} in the intended version {\mycode Jackson-core-asl 1.9.13} or the loaded version {\mycode Jackson-core-asl 1.9.4}, 
a test must instantiate an object of the class {\mycode connector.RESTEasyConnector}
and call the entry method {\mycode createClientExecutor()} provided by the object.
Figure~\ref{fig_2}(c) shows a test generated by \textsc{Riddle} to trigger the invocation of the entry method.
As the randomly generated parameter {\mycode ``a$\ast$\#''} 
is an invalid argument of {\mycode path} to instantiate class {\mycode OpenStackClient}, 
the test will trigger a {\mycode NullPointerException} 
at {\mycode Line 62} of the constructor {\mycode RESTEasyConnector(String)} (as shown in Figure~\ref{fig_2}(a)), 
when it attempts to construct the instance of class {\mycode RESTEasyConnector}. 
As a result, the SC issue is missed. 

%\vspace{-3mm}

%Our observation to address this challenge is that the code snippets of a client project which reference the required object constructors, 
%contain the valid arguments designed by developers for instantiating the corresponding classes.
%For the above example of code snippet, we seed a class instance of class {\mycode Host} with input {\mycode ``AWS-size''} 
%which is extracted from the invocation context of its constructor 
%in source code ({\mycode Line 11} of class {\mycode Host}), 
%to replace the class instance produced by \textsc{Riddle}.
%Figure~\ref{fig_2}(b) describes the generated test, which can precisely captures the different return values of entry method {\mycode Host.onStart()} on these two conflicting library versions, as shown in Figure~\ref{fig_2}(c).
%
%Motivated by the above observations, in this paper, we aim at producing valid class instances during the test generation process, to detect SC issues.
%However, there are no empirical evidences indicate the significance of the valid arguments required by object constructors, in capturing the inconsistent semantic behaviors.
%To achieve our goal, we conducted an empirical study to understand the characteristics of the constructors' arguments designed in the test cases that can expose SC issues.
%Furthermore, we explored whether the constructors with the valid arguments extracted from their invocation contexts of source code can help detect SC issues.
%The above empirical findings can provide guidance for designing our approach.

As observed from the above example, 
the SC issue requires a specific argument to be triggered, i.e., a valid path string {\mycode ``/endpoint''}, which can be found in a caller method of {\mycode RESTEasyConnector}'s constructor (as shown in Figure~\ref{fig_2}(b)).
\textsc{Riddle}, which generates test inputs randomly, is ineffective in detecting such issues. 
%This is because \textsc{Riddle} is essentially a random testing approach. 
%To the best of our knowledge, 
%there are no empirical evidences about
%how significantly the specific test inputs are required in capturing the inconsistent semantic behaviors. 
%the significance of the specific test inputs in capturing the inconsistent semantic behaviors.
The example motivates us to conduct an empirical study to understand 
the characteristics of specific arguments that expose SC issues.

\vspace{-2mm}
\section{Empirical Investigation}
\label{sec:Empirical}

%\wu{Do not mention ``In our approach''. Move the definition of simple constructor and parameterized constructors to Section 2.2}
%In our approach, we refer to an object constructor without inputs as a \emph{simple constructor} and the others that need valid arguments to initiate classes as the \emph{parameterized constructors}.

In this section, we present an empirical study 
on a collection of test cases that have ever successfully captured real SC issues,
with the aim of answering the following two research questions.

%To understand the characteristics of the object constructors that are effective in exposing SC issues, 
%we conducted an empirical study on a collection of test cases that successfully capture the SC issues, with the aim of answering the following two research questions. To ease our presentation, we refer to an object constructor without inputs as a \emph{simple constructor} and the others that need valid arguments to initiate classes as the \emph{parameterized constructors}.

\textbf{RQ1}: \emph{Are randomly generated arguments in test cases likely to capture inconsistent program behaviors of SC issues? What are the characteristics of divergence arguments?}

Generating the desirable objects is a significant challenge for automated test generation techniques~\cite{artzi2008recrash}.
Specifically, 
to increase the likelihood of triggering the conflicting API pairs and revealing SC issues, 
we should be able to generate divergence arguments for object constructors. 
To ease our presentation, we refer to an object constructor taking no arguments as a \emph{no-args constructor} and an object constructor that takes arguments as a \emph{parameterized constructor}. 
In this paper, we focus on the characteristics of the divergence arguments required by parameterized constructors in the test cases, i.e., the concrete values held by arguments, including strings, primitive types, and object references.

%\textbf{RQ2}: \emph{Can the valid arguments of parameterized constructors in the collected test cases be found from the source code of client project?}

\textbf{RQ2}: \emph{Can divergence arguments in test cases be found in the source code of the client project?}

The above investigation can provide empirical evidences and guidance to help construct divergence arguments for parameterized constructors in test generation.

\vspace{-2mm}
\subsection{Collection for benchmark dataset}
\label{sec:data_collection}
%To our knowledge, there is no publicly available test case dataset for detecting real issues caused by inconsistent program behaviors between two code versions.  
%It is difficult to identify the existing tests designed by developers in open source projects, for detecting the real SC issues.
Identifying existing tests written by developers or generated by tools that can detect SC issues is difficult. 
To achieve such a goal, we first simulate a series of dependency conflicts for a given project by altering the actually-loaded versions of its referenced libraries. %\wu{Can we change ``mutate'' into `upgrade''? This would be much better. ``mutate'' seems artificial.}. 
%\scc{upgrading/downgrading = changing? When we say 'actually-loaded', does it mean we have some versions that are loaded but not actually loaded? If not, why don't simply say 'loaded versions'?}
We then execute the project's associated tests to see if it can capture the inconsistent behaviors introduced by the version substitution.
The steps and criteria for constructing such a dataset are described as follows in detail:

\textbf{Step 1: Selecting subjects.} We randomly selected Java projects from {\mycode GitHub} satisfying three conditions: 
(1) including more than 50 test class files %\civi{test methods or test classes?} 
designed by the original developers with their domain knowledge;
(2) passing all the associated test cases without errors (ensuring no SC issues in the selected version); (3) depending on more than 30 libraries (having more upgraded/downgraded candidate libraries). 
As such, we obtained 523 open-source projects.

\textbf{Step 2: Altering the actually-loaded library version.}
%\scc{I prefer: }
For each library on the dependency tree of a subject, we first collected a set of its version numbers released on the {\mycode Maven} central repository, which is denoted as $V = \{v_1, v_2, \cdots, v_n\}$.
We iteratively used each library version $v_i \in V$ to replace its original version on the dependency tree. 
Then, we checked whether the associated tests thrown {\mycode AssertionErrors} when running on the subject after replacements.
The rationale is that the {\mycode AssertionError} in {\mycode JUnit} tests is used to indicate whether the actual variable values are equal to their expected values.
If a test passes for the selected version of a subject and fails for the revised version with an {\mycode AssertionError}, we consider that the failing test captures an SC issue caused by the substitution of library version $v_i$. %\scc{We should use an before SC issue.}
By manually debugging two versions of the program triggered by this failing test, on the execution traces, we can identify a pair of isomerous conflicting APIs defined in both the original and altered library versions.

%For instance, as shown in Figure~\ref{fig_as}, in project {\mycode WZWave 0.0.3}, when we used {\mycode netty-codec} 4.0.33 to replace its original referenced library version {\mycode netty-codec} 4.0.21, its associated test case {\mycode test-ACKPlusMessage()} throwed an {\mycode AssertionError}.
%The reason is that when executing the test case on the mutated version, variable {\mycode out} involved in entry method {\mycode decoder.callDecode(Context, ByteBuf, List)} obtains a different value, compared with the original version.
%Thus, we can conclude that replacing the version of library {\mycode netty-codec} changes the program semantics of client project.

%Two authors were involved in the process of data collection,
Eventually, from 48 Java projects, we obtained 75 SC-revealing test cases, which correspond to 75 conflicting API pairs. %\scc{'code' is uncountable.}
Table~1 shows the statistics of the subjects.
They are large (up to 509.1 kLOC), popular (up to 7,231 stars) and well-maintained (up to 1,108 associated test cases).
Moreover, they have large-scale dependency trees (up to 61 referenced libraries), and on average, each library has 17 versions released on the {\mycode Maven} central repository.
The statistics indicate that the collected test cases are representative.

%\begin{figure}[ht]
%	\vspace{-3mm}
%	\centering
%	\includegraphics[width=0.45\textwidth]{./figure/F_5.pdf}
%	\vspace{-2mm}
%	\caption{An illustrative failing test in our benchmark dataset}	
%	\label{fig_as}
%	\vspace{-4mm}
%\end{figure}

\begin{table*}[]
	\small
	\label{tab}
	\setlength\tabcolsep{6.5pt}     % horizental space
	\def\arraystretch{1.2}
	\caption{The statistics of the subjects collected in our study}
	%	\vspace{-4mm}
	\begin{tabular}{c|c|c|c|c|c|c|c|c|c|c|c|c|c|c|c}
		%	\hline
		\toprule
		\# Project           & \multicolumn{3}{c|}{\# Star} & \multicolumn{3}{c|}{Size(\# kLOC)} & \multicolumn{3}{c|}{\# Test case} & \multicolumn{3}{c|}{\# Library} & \multicolumn{3}{c}{$Avg_v$} \\ \hline
		\multirow{2}{*}{48} & Min.   & Max.    & Avg.   & Min.      & Max.       & Avg.     & Min.      & Max.      & Avg.     & Min.     & Max.     & Avg.     & Min.     & Max.    & Avg.    \\ \cline{2-16} 
		& 73     & 7,231   & 692    & 0.7       & 509.1      & 78.4     & 62        & 1,108     & 201      & 34       & 61       & 40       & 6        & 109     & 17      \\ \bottomrule
	\end{tabular}
	\vspace{-4mm}
\end{table*}

\vspace{-2mm}
\subsection{Empirical findings of RQ1}
\label{sec:RQ1}

%\civi{put such findings at the end of such subsection.}

To answer RQ1, we manually checked 325 class instances used in 75 collected SC-revealing test cases, to analyze the characteristics of the divergence arguments required by their corresponding constructors.
%\civi{why only investigating constructors? we also need valid arguments to invoke APIs?}
By investigation, we found that in the above test cases, 320 out of 325 class instances (98.5\%) need to be created using parameterized constructors, and only five class instances (1.5\%) are constructed without arguments.
%We further investigated the 5 five simple constructors and 
%Note that none of the five simple constructors is the default object constructor.
%\civi{if it does not require any inputs, I think it should be the default object constructor.}
%Although the 5 five simple constructors do not need inputs, the required attributes directly initialized with the assigned values within their method bodies. 
Among 320 parameterized constructors in the test cases, the number of their required arguments ranges from 1 to 7 (\ie 3$\pm$0.34). %\civi{please add the median and the variable}.
Specially, 314 out of 320 parameterized constructors (98.1\%) require more than 2 arguments for creating valid class instances.
For 1,036 divergence arguments needed by the 320 parameterized constructors, we investigated the corresponding source files to understand their characteristics of assignments. Based on our observations, we divide the arguments into the following three types:

%\vspace{-4mm}
\begin{figure}[t!]
	%\vspace{-2mm}
	\centering
	\definecolor{para}{HTML}{FFC0CB}
	\includegraphics[width=0.4\textwidth]{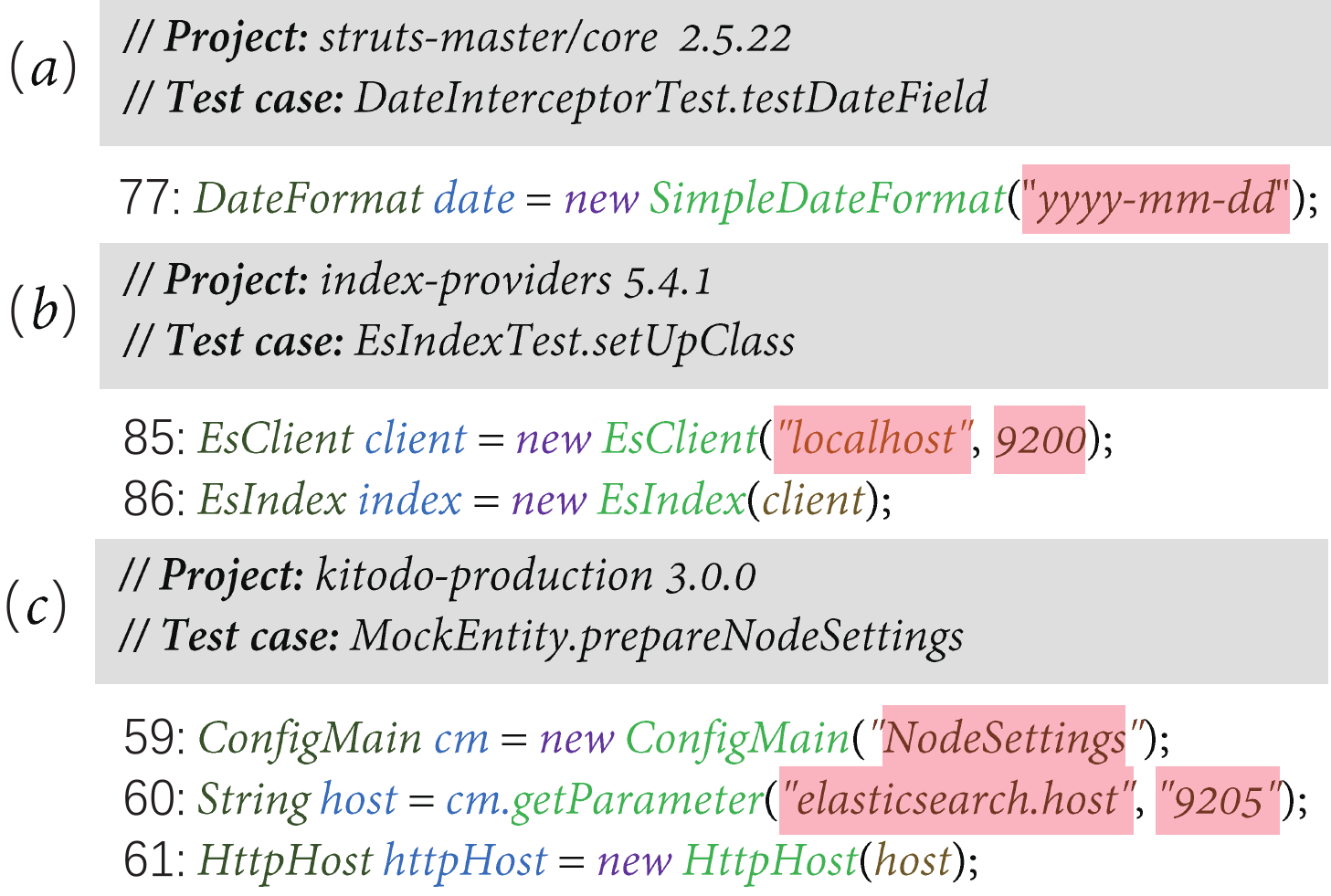}
	%	\vspace{-4mm}
	\caption{The illustrative examples for explaining the constructors' arguments (denoted by \colorbox{para}{arguments}) in test cases}	
	\label{fig_8}
	\vspace{-3mm}
\end{figure}

\textbf{Type 1.} \emph{The arguments are strings or primitive type values with specific semantic meanings or in specific formats (25.3\%).} 262 out of 1,036 arguments are strings or primitive type values (i.e., numeric and enumeration variables) in the test cases. 
By manually checking their values, we found that 168 out of 173 string arguments (97.1\%) are constrained in specific formats or have specific semantic meanings that reflect developers' domain knowledge, such as, the protocol or date related strings.
For instance, in the test file as shown in Figure~\ref{fig_8}(a), a string {\mycode ``yyyy-mm-dd''} is assigned to the parameter of constructor {\mycode SimpleDateFormat(String)}, which can rarely be generated randomly.

Besides, most of the primitive type arguments are used to specify boundary or specific values, e.g., -1, 9200, etc.
For 89 primitive type arguments, we used a random value to replace each argument and then run the corresponding revised test cases on both the original and the altered library versions.
The above process is repeated ten times for each argument.
Over the ten runs, if the revised test cases can capture SC issues once a time, we consider the corresponding argument can be replaced with random values.
%We consider the revised test cases can capture SC issues, if it can 
Unfortunately, 79 out of 89 arguments (88.8\%) failed to trigger the inconsistent behaviors after the random replacements.

%\vspace{-3mm}

\textbf{Type 2.} \emph{The arguments are the instances of other classes created by the constructors with specific inputs (32.4\%).} 336 out of 1,036 arguments are the instances of other classes.
%Among the above 336 arguments, nine (3\%) are {\mycode null} values, 64 (19.0\%) are constructed via static factory methods, 27 (8.0\%) are static fields whose types are class instances, 41 (12.2\%) are constructed by instantiating their corresponding subclasses, and the remaining 195 arguments (58.0\%) are instantiated via accessible constructors.
328 out of the above 336 object constructors (97.6\%) require arguments.
In such cases, developers should recursively construct class instances, and the combination of the involved arguments determines the outermost instance's state.
Therefore, it requires rich domain knowledge to create such combination of specific arguments.
%to create such combination of specific arguments, the requirements for developers' domain knowledge are even higher.
%requires more developers' domain knowledge.
%one's required arguments will affect the others' valid states.
%Such 
%In most cases, the recursively constructed class instances also need specific arguments. 
For example, as shown in Figure~\ref{fig_8}(b), the argument of constructor {\mycode EsIndex(EsClient)} in the test case is created by constructor {\mycode EsClient(String, int)} with specific arguments.
The arguments {\mycode ``localhost''} and {\mycode 9200} determine the state of constructed object {\mycode client}.

\textbf{Type 3.} \emph{The arguments are returned by the other method calls with specific inputs (438/1,036 = 42.3\%).} 
%438 out of the 1036 arguments are the return values of other methods.
In such cases, the states of the constructed class instances are determined by the method calls with valid arguments and the states of class instances providing the above method calls. 
%\civi{I don't understand the above part ``instances offering the above method calls''}.
Figure~\ref{fig_8}(c) shows an example of this type of assignment in test file {\mycode MockEntity.prepareNodeSettings},
the argument of constructor {\mycode HttpHost(String)} is returned by method {\mycode ConfigMain.getParameter(String, String)} with valid inputs {\mycode ``elasticsearch.host''} and {\mycode``9205''}.
In such scenario, developers should also instantiate the classes that provide the required method calls, using specific argument {\mycode``NodeSettings''}.
%the constructors' arguments can hardly be replaced by random values.
%Besides, developers should instantiate the classes offering the required method calls.
Similar to \textbf{Type 2} cases, the combination of the involved arguments required by method calls and the recursively constructed class instances, makes the parameter assignments more complicated.

For most of the parameterized constructors, they require mixed types of arguments.
In many cases, when instantiating a class instance, to construct one type of arguments, we need to recursively create the other types of arguments.
%The recursion depth reflects the complexity of instantiating a class.
%Thus, %it is difficult to assign valid values to the constructors' arguments without the projects' domain knowledge.
An effective technique for the generation of valid class instances is essential to capture real program behaviors.
%To capture real program behaviors, developing an effective technique that can generate valid class instances is necessary.
%Based on the above results, we can distill the following findings:
\vspace{0.6em}
\tcbset{colback=black!10, 
	notitle, 
	width={\linewidth+1pt},
	top=1pt,
	left=1pt,
	right=1pt,
	bottom=1pt,
	toprule=1pt,
	titlerule=1pt,
	bottomrule=1pt,
	leftrule=1pt,
	rightrule=1pt,
	after skip=4pt}

\noindent\begin{tcolorbox}
	\small
	\noindent\textbf{Finding 1:} \emph{320 out of 325 constructors (98.5\%) require arguments to produce valid class instances, in the test cases that successfully capture the inconsistent behaviors.}
	\vspace{0.2em}
	
	\noindent\textbf{Finding 2:} 
	\emph{%We identified three types of arguments required by the 320 parameterized constructors based on their assignment characteristics.
		1,001 out of 1,036 arguments (97.8\%) required by 320 parameterized constructors are subject to semantic constraints, which can hardly be replaced by random values.} %\scc{Unclear to me what semantic meanings refer to. Do we mean the constructor arguments?}
	
\end{tcolorbox}

%\vspace{-4mm}
\subsection{Empirical findings of RQ2}
\label{sec:RQ2}

To answer RQ2, for 320 parameterized constructors in the collected test cases, we first located their corresponding caller methods in the client projects' source code.
Furthermore, we manually checked whether the valid arguments of constructors could be found in the code snippets of these caller methods.
The rational is that the caller methods mostly contain the invocation contexts of a constructor.
Note that an object constructor may have more than one caller methods in the source code.

For each collected parameterized constructor, suppose $Argu_t$ is the total number of the required arguments, and $Argu_s$ is the number of arguments that can be identified in the source code of client project. %, based on the above conditions.
We found that 182 out of 320 parameterized constructors (56.9\%) whose corresponding $Argu_s/ Argu_t$ values are greater than 0.
This means that for 182 parameterized constructors, at least one of their required arguments can be found in the source code.
Specially, among the above 182 parameterized constructors, 119  constructors' corresponding $Argu_s/ Argu_t$ values are equal to 1.
For the rest 63 cases whose corresponding $Argu_s/ Argu_t$ values are between 0 and 1, we observed that 62.2\% of their arguments are passed by the input parameters of the constructors' caller methods.
However, the above caller methods are the APIs provided for invocation by the third-party projects.
As a result, their valid arguments could not be found in the source code of client projects.
%\civi{why we need to check the 63 cases and the following 138 cases separately? Why not just separating into two cases: those arguments that can be found and those that cannot.}

For the remaining 138 parameterized constructors (43.1\%) whose corresponding $Argu_s/ Argu_t$ values are equal to 0, we performed the following tasks:
%\civi{I think the following description to perform the task should be refined.}
(1) We manually extracted the valid arguments from the code snippets of their caller methods. %based on the above five conditions.
If an argument could not be found in the source code, we randomly assigned a value to it.
%In this manner, we constructed a collection of class instances with injected valid arguments;
(2) Let $Argu_s'$ be the number of arguments that are manually extracted from source code.
For each class instances created by the original developers in the test cases, we replaced it with our constructed ones whose corresponding $Argu_s'/ Argu_t$ values are greater than 0 (i.e., at least one argument could be found in the source code).
%We replaced the parameterized constructors created by original developers in the test cases, with our constructed class instances whose corresponding $Argu_s'/ Argu_t$ values are greater than 0;
(3) After the replacement, we executed the revised test cases and checked whether they could still capture the inconsistent behaviors with {\mycode AssertionErrors}.
Finally, 101 out of 138 parameterized constructors (73.2\%) were replaced in 58 test cases.
For the above 58 revised test sripts, 37 of them (63.8\%) successfully detected the SC issues when running on the project versions with upgraded/downgraded libraries.
The average value of $Argu_s'/ Argu_t$ of the 76 replaced constructors in the 37 test cases that can capture SC issues is 0.25 higher than that of the 25 replaced constructors in the 21 test cases that fail to detect SC issues.

%By comparing the $Argu_s'/ Argu_t$ values corresponding to the 76 replaced constructors in 37 test cases that can capture SC issues, and the 25 replaced constructors in 21 test cases that cannot capture SC issues, 
%we found that the on average, the former value is 0.25 higher than that of the latter one.
%Figures~\ref{fig_10}(a) and \ref{fig_10}(b) shows the distribution of the $Argu_s'/ Argu_t$ values corresponding to the 52 replaced constructors in 35 test cases that can capture SC issues and the 37 replaced constructors in 23 test cases that cannot capture SC issues, respectively.
%We can see that most of replaced constructors in Figure~\ref{fig_10}(b) have lower $Argu_s'/ Argu_t$ values than those in Figure~10(a).
%Specially, in the 37 SC-revealing test cases, there are 47 out of 76 replaced constructors (61.8\%), whose $Argu_s'/ Argu_t$ values range from 0.67 to 1.

From the above results, %we can tell that injecting valid arguments
we can draw the conclusion that combining the constructors with their valid arguments extracted from the source code to generate tests can help to expose the inconsistent behaviors. %\scc{It is unclear what exactly invocation contexts refer to?}
The more valid arguments injected to the constructors, the higher success rate of capturing SC issues.
%We can answer RQ2 via distilling the following findings:
%\vspace{-3mm}
\vspace{0.6em}
\tcbset{colback=black!10, 
	notitle, 
	width={\linewidth+1pt},
	top=1pt,
	left=1pt,
	right=1pt,
	bottom=1pt,
	toprule=1pt,
	titlerule=1pt,
	bottomrule=1pt,
	leftrule=1pt,
	rightrule=1pt,
	after skip=4pt}

\noindent\begin{tcolorbox}
	\small
	\noindent\textbf{Finding 3:} \emph{In the collected test cases that can capture the SC issues, 182 out of the 320 parameterized constructors (56.9\%) of which parts of their arguments can be found in the source code.}
	\vspace{0.2em}
	
	\noindent\textbf{Finding 4:} \emph{When we substituted our injected arguments for the constructor arguments that cannot be found in the source code, %into which we manually injected valid arguments from source code, 
		37 out of 58 test cases (63.8\%) captured SC issues.}
	%\vspace{0.2em}
\end{tcolorbox}
\vspace{0.4em}
%From the discussions, we can draw the conclusion that combining the constructors with their invocation contexts extracted from the source code to generate tests can help to capture the inconsistent program semantics.
The empirical findings of RQ1 and RQ2 shed lights on understanding the characteristics of the object constructors' arguments and provide valuable guidance to design an automated test generation technique for detecting SC issues.

\vspace{-1mm}

\section{Methodology}
\label{sec:Methodology}
\vspace{-1mm}
\subsection{Sensor in a nutshell}
\label{sec:Overview}
\begin{figure*}[t]
	\centering
	\includegraphics[width=0.8\textwidth]{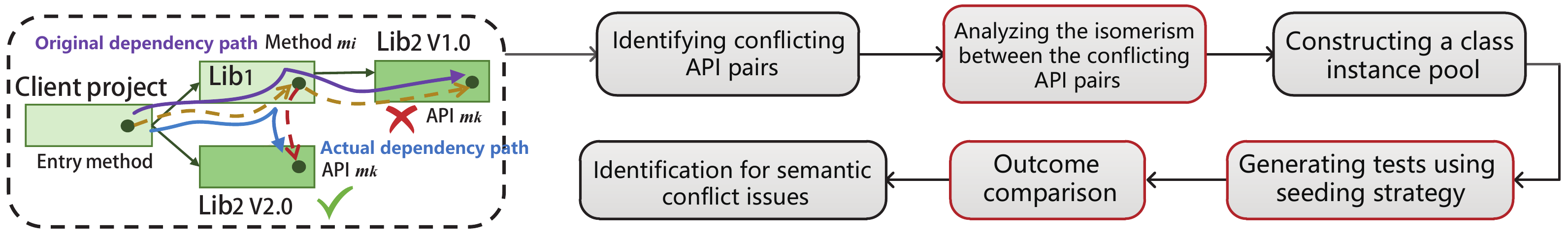}
	%	\vspace{-4mm}
	\caption{The overall architecture of \textsc{Sensor}}	
	\label{fig_o}
	%	\vspace{-5mm}
\end{figure*}

Figure~\ref{fig_o} shows an overview of our approach, which involves three steps: \emph{identification for isomerous conflicting API pairs}, \emph{test generation} and \emph{outcome comparison}. %\scc{Have we explain 'isomerous' in preceding paragraphs? It seems that we explain it in Section 4.2.}
%\civi{Can we highlight the major steps in Figure~\ref{fig_o}?}.
First, \textsc{Sensor} finds a set of conflicting API pairs introduced by a dependency conflict and identifies which of these conflicting API pairs are isomerous. %\scc{I think it is better to introduce isomerous conflicting API when discussing empirical observation in Section 3.}
%Based on the \textsc{Gumtree} technique~\cite{falleri2014fine}, it identifies the isomerous conflicting API pairs.
Second, it generates tests to capture the inconsistent variable states of a given client project affected by isomerous conflicting API pairs.
Third, by comparing their test outcomes obtained on two conflicting class versions, \textsc{Sensor} identifies the SC issues and points out their impacts on the client project's program behaviors at a fine-grained level to help diagnose SC issues.
%at a fine-grained level by pinpointing which states of variables involved in the invocation contexts  differ in which test,

%It mainly contains the following three steps:

%\vspace{-5mm}

\vspace{-3mm}
\subsection{Identifying isomerous conflicting API pairs}
\label{sec:Static analysis}
%In static analysis, 
%\textsc{Sensor} first identifies a set of conflicting API pairs introduced by each dependency conflict.
\textsc{Sensor} identifies the isomerous conflicting API pairs at a fine-grained level based on code differences detected iteratively using \textsc{Gumtree}~\cite{falleri2014fine}.
%with the aim to prioritize based on their impacts on the client project's semantics. 
It considers that those conflicting API pairs with different implementations will potentially cause semantic conflicts.

%\civi{we need to explain why we need to measure the isomerism.}

\textbf{Identifying conflicting API pairs.} 
%Given a project, \textsc{Sensor} extracts the library dependency tree from its dependency management script (e.g., {\mycode pom.xml}).
By analyzing the dependency tree of a client project, \textsc{Sensor} identifies multiple versions of a class $C_i$ or a library $Lib_i$.
%In this paper, \textsc{Sensor} identifies the shadowed class version or library version based on {\mycode Maven}'s class loading mechanism, which has been systematically summarized by a recent study~\cite{wang2018dependency}.
For each API $C_i'\!::\!m_k$
% \civi{symbols needs to be changed. $Lib_{sd_{it}}$ has never been introduced before.} 
defined in the shadowed class version $C_i'$ and referenced by the client project, \textsc{Sensor} considers the API $C_i\!::\!m_k$ that satisfies one of the following two conditions as its replaceable method:
%\civi{What if there are multiple replaceable methods for a method?}
\begin{itemize}[leftmargin=*]
	\item API $C_i\!::\!m_k$ has the same signature (i.e, method name, parameter types and return types) as $C_i'\!::\!m_k$ and is defined in the actually-loaded class version.
	\item Suppose class $CS_{{i}}$ is the superclass of $C_{i}$. If the actually-loaded class $C_i$ does not include the API with the same signature as $C_i'\!::\!m_k$, then \textsc{Sensor} regards the API defined in its superclass $CS_{i}\!::\!m_k$ that can be overridden by $C_i'\!::\!m_k$ as its replaceable method.
	In this case, the API compatibility will not be broken due to dynamic binding mechanisms.
\end{itemize}

Finally, $C_i'\!::\!m_k$ and $C_i\!::\!m_k$  
%\civi{I think it is possibe that $m_k(Lib_{sd_{it}})$ is a set of methods}
are identified as a pair of conflicting APIs, which is denoted as $RP\langle C_i^\prime\!::\!m_k, C_i\!::\!m_k\rangle$.

%\ying{Rongxin, could you give some comments on the new strategy to identify isomerism conflicting API pair combining Gumtree and call graph?}

\textbf{Analyzing isomerous conflicting API pairs.}
We adopt \textsc{Gumtree} \cite{falleri2014fine} to check if implementation differences exist in a conflicting API pair.
%to identify the fine-grained code differences between a conflicting API pair.
\textsc{Gumtree} detects code differences based on abstract syntax trees (ASTs).
%Specifically, it generates four types of differences, including adding, deleting, updating, and moving nodes, at the AST level. 
When applying \textsc{Gumtree}, \textsc{Sensor} needs to consider the cases where a pair of conflicting APIs exhibit no difference in terms of their ASTs 
but are semantically different due to the changes in their depended methods. 
%However, even though the a pair of conflicting APIs exhibit no difference in terms of their corresponding ASTs, 
%their semantics might still be different if the methods invoked by these two APIs have been changed.
Typically, an API could invoke a series of methods which constitute a call graph.
Any changes in the methods invoked by an API could possibly affect the states of its referenced variables, thereby changing the API's semantics.
To perform a comprehensive analysis for a conflicting API pair, we construct the corresponding call graphs of the two APIs, and then iteratively compare each method pair with the same signature on the call graphs in a top-down manner.
In the process of iterative analysis, we consider the above conflicting API pair as an \emph{isomerous conflicting API pair}, if there are AST differences identified by \textsc{Gumtree} between one comparable method pair on the call graphs.

Although our iterative analysis can capture all code differences between a pair of method invocation paths on the call graphs, not all the differences are useful in practice. 
According to an empirical study conducted by Schr{\"o}ter et. al \cite{schroter2010stack}, nearly 90\% of the issues are fixed within the \textbf{top-10} methods along the invocation paths.
Other deeper methods barely affect the program semantics of the client project.
To reduce such false positives, \textsc{Sensor} only analyzes the methods whose call depth is less than \textbf{ten}, along the original and actual dependency paths of a conflicting API pair.

\vspace{-4mm}
\subsection{Test generation}
\label{sec:Test generation}
\textsc{Sensor} is built on top of \textsc{Evosuite}, which adopts a genetic algorithm (GA) to derive a test suite for a given target class.
A target class is the one that contains an entry method that directly or indirectly references the identified isomerous conflicting APIs along their original and actual dependency paths, respectively.
%A test case is essentially a program that executes the SUT, which consists of a sequence of statements,
%where each statement in a test case can generate class instances through constructors and can access fields and methods offered by these instances.
%The GA works by iteratively selecting individuals from the population based on their fitness with respect to the search objective.
%Note that the individuals of the GA's population are test suites.
\textsc{Sensor} adopts the fitness function defined by \textsc{Riddle}, which aims to maximize the possibility of covering the identified isomerous conflicting APIs~\cite{WANG2019STACK}.

To precisely capture the program behaviors, %\textsc{Sensor} adopts a seeding strategy for creating class instances.
%Existing techniques~\cite{alshahwan2011automated, rojas2016seeding, fraser2012seed, sakti2015instance, mcminn2012search, alshraideh2006search} have proposed different seeding strategies for test input generation, especially for strings and primitive types.
%They  the values of these strings and primitive types designed in the source code when generating tests.
%\civi{It seems that our approach also leverages those values designed in the source code? what is the difference?}
%However, these techniques cannot map the collected values into the parameters of the method calls and object constructors, as they do not consider their corresponding invocation contexts.
%Since these strategies generate constructor arguments without considering their invocation contexts, they are not effective in constructing valid class instances to expose conflicting API pairs.
\textsc{Sensor} adopts a seeding strategy of class instances inspired by the our empirical findings summarized in Section~\ref{sec:Empirical}.
%Besides the seeding strategy as adopted by \cite{sakti2015instance}, 
\textsc{Sensor} injects the invocation context information extracted from the source code into class instances with the aim of generating divergence arguments.
%\civi{I am not sure whether ``variable states'' is a good term since ``states'' always gives the impression of dynamic execution.}
%We observe that the code snippets of source files contain the \emph{valid arguments} designed by developers using their domain knowledge for instantiating the corresponding classes.
Specifically, \textsc{Sensor} first constructs a pool of instances with the injected invocation contexts for each class included in a client project, which is denoted as $CIP$.
When \textsc{Evosuite} needs to instantiate a class in a test, \textsc{Sensor} tries to select an instance from $CIP$ and provide it to \textsc{EvoSuite}. %\ying{I should claim the novelty of the seeding strategy, compared with the existing seeding approaches.}

\textbf{$CIP$ Construction.} %Inspired by Sakti et al.'s approach~\cite{sakti2015instance}, we use the term \emph{means-of-instantiation} to represent any means of instantiating a given class offered in a program.
\textsc{Sensor} constructs a class instance involving two steps: identifying its possible object constructors and extracting the constructors' invocation contexts from the source code of the client project.
%For a given class $C_i$, \textsc{Sensor} considers four types of object constructors, Sakti et al.'s approach~\cite{sakti2015instance}: % which are observed in our empirical study in Section~\ref{sec:Empirical}:
%(1) all accessible constructors;
%(2) all static factory methods returning an instance of class $C_i$;
%(3) all static fields that are instances of class $C_i$;
%(4) all external methods that returns an instance of class $C_i$, i.e., methods that return an instance of a required class and are defined outside of that class;
%and (4) the above methods that can instantiate the subclasses of class $C_i$, which can also be used to instantiate abstract classes.
Let $MOI_i = \{mo_{1}, mo_{2}, \cdots, mo_n\}$ be a set of possible object constructors of a given class $C_i$ collected by
the static analysis approach \cite{sakti2015instance},
% \textsc{Sensor} using static analysis (i.e., using the {\mycode soot}~\cite{Soot} framework) based on approach~\cite{sakti2015instance}, 
where $mo_{k} \in MOI_i$ represents an object constructor of this class.
For most of the cases, a constructor requires parameters for substantiation.
To inject valid arguments into $\forall mo_{k} \in MOI_i$, \textsc{Sensor} performs the following tasks: 
(1) it identifies a set of caller methods $M_k = \{m_{1}, m_{2}, \cdots, m_n\}$ in the client project, which reference the constructor $mo_k \in MOI_k$;
and (2) for each caller method $m_{t} \in M_k$, \textsc{Sensor} locates the source file where it is defined, and considers the code snippets within the source file as a search scope of the invocation contexts of $mo_{k}$.
Specially, \textsc{Sensor} takes into account the following three cases to extract the invocation contexts of $mo_{k} \in MOI_i$, based on three types of constructors' arguments observed in our empirical study:
%\civi{I am still not quite understand what is ``invocation contexts''.}
%For instance, as shown in Figure~\ref{seed_updated}, \textsc{Sensor} can find four caller methods of constructor {\mycode AttributeFilter(String, String)} in project {\mycode FluentLenium-3.9.1}, i.e., {\mycode filterAttributeIsIDA()},  {\mycode filter-\\AttributeIsTextContent()}, etc.
%Then, in the source files where the identified caller methods defined (e.g., {\mycode core.filter.Attribu-\\te.java}), 

%\begin{itemize}[leftmargin=*]
\emph{Case 1:} If the required arguments of $mo_{k}$ are strings or primitive types that can be identified in source code by exactly matching assignment statements with variable names, \textsc{Sensor} extracts their corresponding assigned values directly from the source code.
%For instance, as shown in Figure~\ref{seed_updated}, in file {\mycode core.filter.Attribute.java}, \textsc{Sensor} directly extracts arguments of {\mycode AttributeFilter(String, String)} by mapping its variable names with the assignment statements in {\mycode Lines} 257 and 258.

\emph{Case 2:} In the case where the arguments of $mo_{k}$ are the input arguments of the caller method $m_i \in M_k$, \textsc{Sensor} recursively searches the corresponding invocation context for caller method $m_i$.
%For example, in file {\mycode core.filter.FilterConstructor.java}, a required argument of constructor {\mycode AttributeFilter(String, String)} is assigned by the input parameter of its caller method {\mycode withId(String)}.
%In this scenario, \textsc{Sensor} recursively searches its valid argument in the invocation context of {\mycode withId(String)} defined in other files.

\emph{Case 3:} For the arguments whose assigned values are returned by the other method calls, \textsc{Sensor} %considers the method calls as the invocation contexts of $mo_{k}$.
%Furthermore, it 
recursively finds invocation contexts of the required method calls and the class instances that provide the method calls, following the above steps.

For the required arguments that are the instances of other classes, \textsc{Sensor} recursively constructs such class instances following the above steps.
Moreover, if the required arguments are strings or primitive types whose valid values cannot be exactly extracted from source code, \textsc{Sensor} randomly assigns values to them.
In the cases of recursively constructing class instances or searching for valid arguments from the intermediate invocation contexts (e.g., \emph{Cases 2} and \emph{3}), the searching process is terminated if the recursion depth is greater than $DN$, or it cannot generate an instance of the current class (e.g., the class is not instantiable for accessibility reasons).
%Note that, when the searching process is terminated 
%\ying{I should use the cases to explain the seeded context.}
%In our approach, we set $DN = 5$ empirically.
In this manner, for each class $C_i$, we can obtain a set of possible object constructors $MOI_i$, and each constructor $mo_{k} \in MOI_i$ corresponds to a set of invocation contexts extracted from the source code. %\civi{What are ``invocation contexts''?}

\textbf{Seeding strategy.} %To integrate the $CIP$ into test generation, \textsc{Sensor} further defines a new type of statement (\emph{class instance statement}) in \textsc{Evosuite}, which represents the creation of a specific class instance.
During the insertion of new statements into a test case, %(when searching or initializing the population), 
\textsc{Evosuite} tries to resolve dependencies either by re-using objects declared in earlier statements of the same test, or by recursively inserting new calls to generate new instances of the required dependency objects~\cite{fraser2016evosuite, arcuri2014automated}.
Whenever \textsc{Evosuite} attempts to generate a class instance, \textsc{Sensor} selects one of the instances of that class from $CIP$ with probability $P_{OC}$ to replace it.
%\civi{where does this probability comes from?}
Note that if $CIP$ does not contain the required class instance, such replacement operations are not performed.

In the presence of different object constructors and their various corresponding invocation contexts for a given class, setting the probability $P_{OC}$ to choose one of them is challenging.
%For instance, in project {\mycode MiA}~\cite{MiA}, most classes have more than 20 \emph{means-of-instantiations} with more than 50 invocation contexts on average (e.g., class {\mycode Train.TrainNewsGroups} corresponds 58 different constructed class instances in $CIP$).
In our approach, we set a probability $P_{OC}$ based on the complexity of a constructor's invocation contexts.
The selection strategy favors the object constructor with a low complexity.
Also, it dynamically adjusts the probability $P_{OC}$ according to the number of times that a class instance has been selected during the test generation process, to diversify candidate class instances.
Thus, we have
\vspace{-1mm}
\begin{equation}
\label{eqn_1}
\small
P_{OC} = %\frac{1}{Depth_{arg}} \times \frac{1}{T_{s}}
1 /{Depth_{arg}} \times 1 / {T_{s}}
%\dfrac{1}{3}
\vspace{-1mm}
\end{equation}
where $Depth_{arg}$ is an indicator of a constructor's complexity, which represents the recursion depth for constructing the involved class instances or searching valid arguments from the intermediate invocation contexts; 
and $T_{s}$ is the number of times that a class instance has been seeded into the tests.
%Based on Equation~(\ref{eqn_1}), \textsc{Sensor} balances the diversity and complexity of our seeding strategy.
%Sometimes, 
%succeeds to be seeded into a test case without causing test failures.
%The detailed selection strategy can refer to approach~\cite{sakti2015instance}.%\ying {I should explain how we set the probability $P_{OC}$.}

\vspace{-2mm}
\subsection{Test outcome comparison}
\label{sec:Issue prioritization}
%Based on the proposed test generation technique, \textsc{Sensor} compares the differences of the client project in terms of their semantic behaviors when the executions of any pair of isomerous conflicting APIs are triggered.
%In our approach, \textsc{Sensor} does not consider that all the inconsistent semantic behaviors indicate bugs.
% \civi{behavioral differences are not necessarily bugs or just we do not consider them as bugs?}
%It concretely exposes behavioral differences at a fine-grained level by pointing out the inconsistencies in variable states.
%\textsc{Sensor} supports the comparison at two levels, i.e., intermediate states of a program and test outcomes, and assigns a score to each isomerous conflicting API pair under test according to their comparison results.
%For each isomerous conflicting API pair $RP\langle C_i^\prime\!::\!m_k, C_i\!::\!m_k\rangle$, \textsc{Sensor} identifies a set of entry methods $EM$ defined in the classes of the client project that indirectly references API $C_i'\!::\!m_k$.
%It then generates tests for the classes in which the above entry methods define, 
For each isomerous conflicting API pair, \textsc{Sensor} generates tests to trigger their executions on the actually-loaded and the shadowed class versions where they are defined, respectively.
It repeats the above test generation process for $RN$ times and then compares the test outcomes.
\textsc{Sensor} takes the following two types of semantic inconsistencies into account, in the comparison process:
\begin{itemize}[leftmargin=*]
	\item \emph{Variable states.}
	It considers three types of affected variables, including: (a) each input parameter of entry method whose type is an object; (b) each variable used by entry method but not defined in it; and (c) return variable of the entry method.
	As a result, if the state of any affected variable is different across the executions on the above two versions of code, \textsc{Sensor} regards the behaviors as inconsistent.
	\item \emph{Test outcomes.} If a test succeeds to run on one version of code but fails on the other, a semantic inconsistency is noted. 
\end{itemize}

\textsc{Sensor} considers the isomerous conflicting APIs that induce inconsistent behaviors when executing more than one generated tests, as the cases that could cause SC issues.
Finally, it illustrates their impacts on the program semantics of the client project, to help developers further diagnose the SC issues.

\vspace{-2mm}
\section{Evaluation}
\label{sec:Evaluation}
This section presents our experimental results through answering the following research questions:
%Our tool and experimental data are publicly available at \underline{https://sensordc.github.io/}.
\begin{itemize}[leftmargin=*]
	\item \textbf{RQ1 (Effectiveness):} How effective is \textsc{Sensor} in detecting SC issues? 
	\item \textbf{RQ2 (Usefulness):} Can \textsc{Sensor} detect unknown SC issues and provide useful diagnosis information? 
\end{itemize}

\vspace{-2mm}
\subsection{Experimental design}

\subsubsection{RQ1}
To study RQ1, we first collected a high quality ground truth dataset and then applied \textsc{Sensor} to this dataset to assess its effectiveness in detecting SC issues.

\noindent\textbf{Collection for the ground truth dataset.} We consider 75 isomerous conflicting API pairs that can cause {\mycode AssertionErrors} when executing 75 tests collected in our empirical study (in Section~\ref{sec:Empirical}), as the ones introducing SC issues into their client projects.
We labeled the isomerous conflicting API pairs that will not cause SC issues in client projects, based on the following steps:

(1) We mined the historical commits of open-source projects on {\mycode Github} and identified the commits that only upgraded/downgraded a library version in the projects' dependency management scripts (e.g., {\mycode pom.xml}).
We consider that the above change of a library version does not affect the client project's program behaviors, if its corresponding commit satisfies all the following conditions: 
\begin{itemize}[leftmargin=*]
	\item All the tests triggered by a continuous integration build tool (e.g., {\mycode TravisCI}) can pass the revised project version, after this commit is submitted.
	
	\item In the case that the client project is still active, there are no version changes for this library, in the next 24 months after the above commit being merged.
	Besides, during the above period, there are no issues or commits whose descriptions and logs mentioning the semantic issues caused by this revised library version.
	The rational is that a recent study~\cite{kim2006long} found that bugs are usually repaired within 2 years across different projects since they were introduced into the project.
	%Note that we consider a client project is active, if it has commit records within three months.
\end{itemize}

(2) For an identified semantic-preserving library version change, we labeled the isomerous conflicting API pairs in the original and revised versions of this library, which can be covered by the client project's tests (triggered by the continuous integration build tool) 
without errors, as the ones that will not introduce SC issues.

Eventually, we collected 150 isomerous conflicting API pairs that will not introduce SC issues into their client projects and 75 isomerous conflicting API pairs definitely causing SC issues.
%The data collection process involves two graduate students: one identifying the 

\noindent\textbf{Metrics.} 
The outcomes of \textsc{Sensor} can be categorized as follows:
%which are defined as follows:
(1) \textit{True Positive} (TP): The inconsistent behavior identified by \textsc{Sensor} between a conflicting API pair is a real SC issue.
(2) \textit{False Positive} (FP): The inconsistent behavior identified by \textsc{Sensor} between a conflicting API pair is not a real SC issue.
(3) \textit{True Negative} (TN): No inconsistent behavior is identified by \textsc{Sensor} between a conflicting API pair, and it is not a real SC issue.
(4) \textit{False Negative} (FN): No inconsistent behavior is identified by \textsc{Sensor} between a conflicting API pair, but it is a real SC issue.
Based on the outcomes, we use \textit{Recall}, \textit{Precision}, and \textit{F-measure} to evaluate the performance of \textsc{Sensor}, which are defined as follows.  
%The evaluation metrics are defined as follows:
%Based on the above outcomes, we obtain:
\vspace{-2mm}
\begin{equation}\label{eqn_precision}
{Precision = }{{TP} \mathord{\left/
		{\vphantom {{TP} {TP + FP}}} \right.
		\kern-\nulldelimiterspace} {(TP + FP)}}
\vspace{-3mm}
\end{equation}

\vspace{-1mm}
\begin{equation}\label{eqn_recall}
{Recall = }{{TP} \mathord{\left/
		{\vphantom {{TP} {(TP + FN)}}} \right.
		\kern-\nulldelimiterspace} {(TP + FN)}}
\vspace{-3mm}
\end{equation}
\vspace{-2mm}
\begin{equation}\label{eqn_fmeasure}
{F\text{-}measure = }{{2 \times {Precision} \times {Recall}} \mathord{\left/
		{\vphantom {{2 \times {Precision} \times {Recall}} {{Precision} + {Recall}}}} \right.
		\kern-\nulldelimiterspace} {({Precision} + {Recall})}}
\vspace{-1mm}
\end{equation}

\textit{Precision} evaluates whether \textsc{Sensor} can detect SC issues precisely. 
\textit{Recall} evaluates the capability of \textsc{Sensor} in detecting all the SC issues.
$F$-$measure$ takes the $Precision$ and $Recall$ into consideration, and weights these two metrics equally~\cite{wu2011relink}.

\noindent\textbf{Comparison.} We compared \textsc{Sensor} with the latest dependency conflict detection tool \textsc{Riddle}, in terms of their effectiveness in covering the target branches.
\textsc{Riddle} is chosen as the baseline because it is designed for generating tests to trigger the program execution from an entry method of client project to reach an identified conflicting API that is missing in the actually-loaded library, thereby causing a program crash.
We consider that \textsc{Riddle} detects a SC issue, if its generated tests can trigger the isomerous conflicting API pairs in the ground truth dataset and capture the variable state or test outcome inconsistencies.

\noindent\textbf{Experimental setting.}
For both \textsc{Sensor} and \textsc{Riddle}, we set the time budget for the evolutionary search to 800 seconds and repeated the test process for $RN = 10$ times on each code version with different random seeds.
The final results were averaged over the ten runs to avoid the biased results.
%In the cases that recursively constructing class instances or searching divergence arguments for the intermediate invocation contexts, %\textsc{Sensor} terminates the searching process if the recursion depth is greater than $DN = 5$.
%\wu{Can we skip this detail? Otherwise, we should consider to add the justifications on why we set DN as 5. }

%\vspace{-2mm}
\subsubsection{RQ2}
%To answer RQ2, (detect real SC issues and submit issue reports)
To answer RQ2, we conducted experiments on 92 open-source Java projects randomly sampled from {\mycode GitHub} using three criteria: (1) it has received more than 50 stars or forks (\ie popularity); (2) it references multiple versions of libraries or classes detected by static analysis and contains at least one commit after December 2019 (i.e., actively-maintained); (3) it is not included in the subject set of our empirical study in Section~\ref{sec:Empirical}  (i.e., new validation).
We leveraged \textsc{Sensor} to generate issue reports that include:
(1) the root causes of SC issues;
(2) the isomerous conflicting API pairs that induce semantic inconsistencies and their corresponding original and actual dependency paths;
(3) the generated test cases that can trigger the executions of the isomerous conflicting API pairs; 
(4) the differences in the test outcomes or variable states of the client project after executing the generated test cases.
We submitted the issue reports to the corresponding developers via the projects' issue tracking systems and evaluated the usefulness of \textsc{Sensor} based on developers' feedback.
%To avoid overwhelming developers with all the isomerous conflicting API pairs and the diagnostic information, we reported the one with the highest score in detail, since the semantic changes that it induces to the host project are witnessed by the largest number of tests.
%Besides, we also provided the other isomerous conflicting API pairs' information as an attachment of the issue report. 

\vspace{-1mm}
\subsection{RQ1: Effectiveness of Sensor}
\label{sec:Effectiveness}
\subsubsection{Overall effectiveness.}

Table 2 shows the experimental results on the ground truth dataset.
\textsc{Riddle} identified 8 SC issues with 2 (25.0\%) false positives.
Besides, it did not capture any inconsistent behaviors between 217 isomerous conflicting API pairs, with 69 (31.8\%) false negatives.
Our approach, \textsc{Sensor}, identified 71 SC issues with 14 (19.7\%) false positives, which achieves a $Precision$ of 0.803.
For the isomerous conflicting API pairs that will not cause SC issues in client projects, \textsc{Sensor} successfully identified 154 of them, with 18 (11.7\%) false negatives, leading to a $Recall$ of 0.760.
In terms of the $F$-$measure$, \textsc{Sensor} also significantly outperformed \textsc{Riddle} (0.781 vs. 0.145).

By manually checking the six true positive SC issues detected by \textsc{Riddle}, we found that in these cases, all the invocation depths from the entry methods of client projects to the conflicting APIs are less than 3.
All the object constructors in the test cases generated by \textsc{Riddle} that successfully captured the SC issues, do not need arguments.
However, the initialization of the required variables could be found in the method bodies of the above specific simple constructors.
Therefore, they could easily trigger the target branches with valid program semantics.
Note that the above six true positive SC issues were also detected by our technique.
For 57 true positive cases detected by \textsc{Sensor}, the average invocation depth from the entry methods to the conflicting APIs is 6.9.
It is largely ascribed to the effectiveness of \textsc{Sensor}'s seeding strategy of class instances.
The seeded objects greatly increase the possibility of reaching the target branches, compared with \textsc{Riddle}.
We further investigated the reasons why \textsc{Sensor} generated false positive and negative cases of SC issues and summarized them below.
\begin{table}[t!]
	%	\small
	\footnotesize
	%	\vspace{-3mm}
	\caption{The experimental results on the ground truth dataset}
	\bgroup
	\setlength\tabcolsep{4.1pt}     % horizental space
	\def\arraystretch{1.2}
	%	\vspace{-4mm}
	\begin{tabular}{c|c|c|c|c|c|c|c}
		%	\hline
		\toprule
		& $TP$ & $FP$ & $TN$  & $FN$ & $Precision$ & $Recall$ & $F$-$measure$ \\ \hline
		\textsc{Sensor} & 57 & 14 & 136 & 18 & 0.803    & 0.760 & 0.781     \\ \hline
		\textsc{Riddle} & 6  & 2  & 148 & 69 & 0.750    & 0.080  & 0.145     \\ \bottomrule
	\end{tabular}
	\egroup
	%	\vspace{-3mm}
\end{table}

\begin{figure}[t!]
	%\vspace{-3mm}
	\centering
	\includegraphics[width=0.4\textwidth]{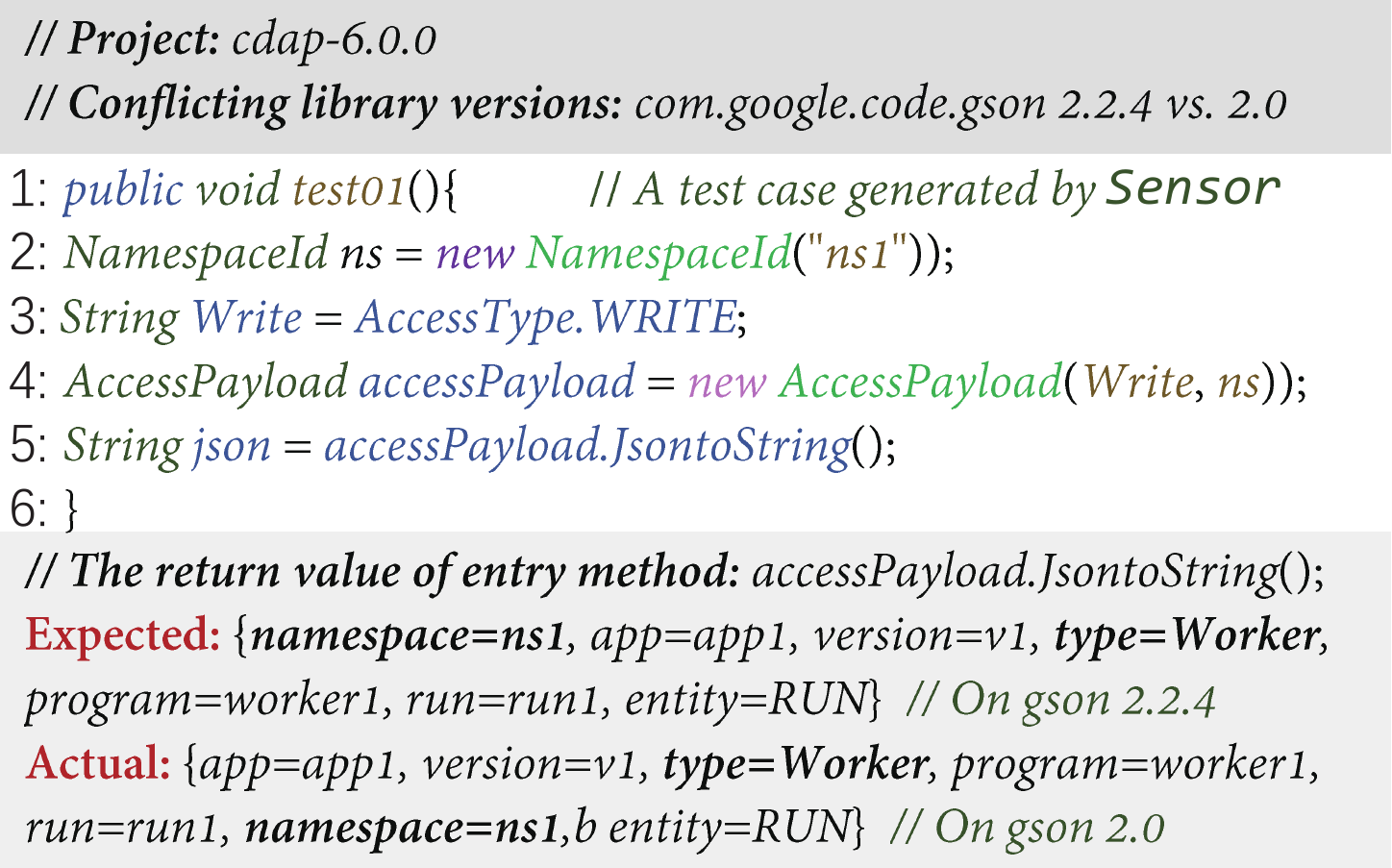}
	%	\vspace{-4mm}
	\caption{A false positive example}	
	\label{fig_fp}
	%	\vspace{-4mm}
\end{figure}

\noindent\textbf{False positive examples.} The main cause of false positives generated by \textsc{Sensor} is that the inconsistent behaviors are caused by the non-deterministic or random variable states, which are benign for the client projects.
For example, as shown in Figure~\ref{fig_fp}, a test case generated by \textsc{Sensor} captured the inconsistent return values of the entry method in project {\mycode cdap 6.0.0}, on conflicting versions of library {\mycode com.google.code.gson}.
The return value is a {\mycode Json} string, which has the same attributes but different declaration orders on these two library versions.
However, the attributes are stored in an unordered collection and the sequence of traversing the attributes is non-deterministic in the program.
Such differences do not affect the semantics of client project, and therefore it did not catch developers' attention.
The other false positives detected by both \textsc{Sensor} and \textsc{Riddle} are similar cases, in which the inconsistent behaviors affected by conflicting API pairs (e.g, non-deterministic text formats, random values, etc.), are benign for the program semantics.

%\vspace{-2mm}

\noindent\textbf{False negative examples.} We manually investigated the 18 false negative cases detected by \textsc{Sensor}, and divided them into the following two categories: 
\begin{itemize}[leftmargin=*]
	\item \emph{The inconsistent branches within conflicting API pairs cannot be reached} (10/18 = 55.6\%).
	In these cases, the required object constructors have multiple caller methods with different invocation contexts in the source code, and the ones seeded by {\mycode Sensor} could not trigger the inconsistent branches within conflicting API pairs, even over the ten runs.
	%	For instance, as shown in Figure~\ref{fig_fn1}, an inconsistent branch in API {\mycode Base64.decodeBase64(byte[])} could be triggered when the length of variable {\mycode base64Data} is greater than or equal to 4.
	%	And the value of {\mycode base64Data} is indirectly passed by the input of constructor {\mycode EdmBinay(String)}.
	%	However, this constructor has seven caller methods with different invocation contexts in the source code.
	%	In the process of generating tests, \textsc{Sensor} did not select the argument that can exactly reach the concerned branch based on the probability we defined in Equation~(\ref{eqn_1}), as its corresponding recursion depth for searching the intermediate invocation contexts is relatively great (five).
	\begin{comment}	
	\begin{figure}[ht]
	%	\vspace{-2mm}
	\centering
	\includegraphics[width=0.4\textwidth]{./figure/F_10.pdf}
	%	\vspace{-3mm}
	\caption{A false negative example}	
	\label{fig_fn1}
	\vspace{-4mm}
	\end{figure}
	\end{comment}	
	
	\item \emph{The program crashed before reaching the conflicting API pairs} (8/18 = 44.4\%).
	The eight false negative cases were caused by the inaccurate arguments extracted by \textsc{Sensor}, which led to program crashes before triggering the conflicting API pairs.
	By further investigation, 
	we found that the inaccurate arguments are manifested into two patterns:
	(1) \emph{the required arguments are affected by a series of method calls that cannot be extracted from the source code exactly} (5/8 = 62.5\%), and (2) \emph{part of a constructor's arguments cannot be found from the source code, to which random values are assigned} (3/8 = 37.5\%).
\end{itemize}

%\begin{comment}

\vspace{1mm}
\subsubsection{Effectiveness on producing valid class instances.}
%To quantitatively evaluate the effectiveness of \textsc{Sensor} on producing valid class instances, we further calculate the following measurements:

%\begin{itemize}[leftmargin=*]
%	\item Let $N_c$ be the number of classes in a project for which \textsc{Sensor} could construct instances and $N_t$ be the total number of classes in this project. 
%	Then, $N_c/N_t$ describes the \textsc{Sensor}'s capability on constructing class instances with extracted invocation contexts.

%	\item Let $N_i$ be the average number of instances with different divergence arguments constructed for each class in a project.
%The more diverse instances for a class are constructed, the higher probability of capturing inconsistent semantic behaviors in different invocation contexts would be.

%	\item Let $Argu_s$ be the number of arguments required by a constructor that can be extracted from the source code and $Argu_t$ be the total number of required arguments in this constructor.
%	Then, $Argu_s/ Argu_t$ indicates the correctness of the state of any class instance constructed in $CIP$.

%	\item Let $Rp$ represent the number of class instances that are successfully seeded by \textsc{Sensor} in a test case and $N_o$ be the total number of instantiated classes in this test case.
%	Then, $Rp/ N_o$ is the substitution rate of class instances by our approach in a test case. 

\begin{figure}[!t]
	\vspace{-3mm}
	\centering	\includegraphics[width=0.45\textwidth]{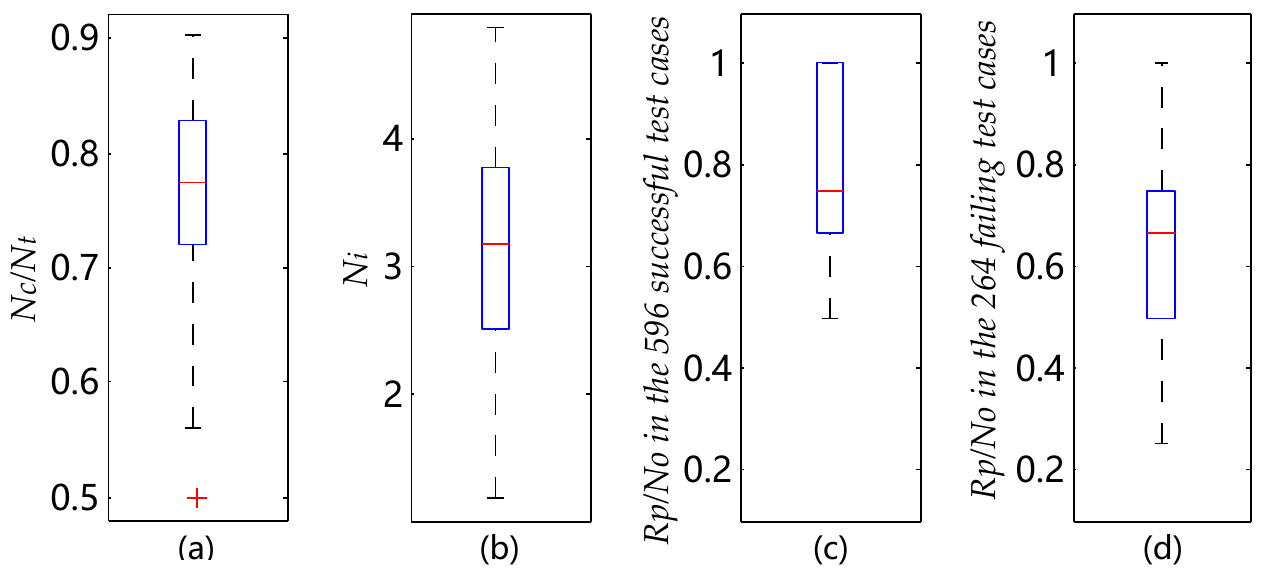}
	%\vspace{-4mm}
	\caption{Effectiveness on producing valid class instances}	
	\label{fig_122}
	%\vspace{-3mm}
\end{figure}

Let $N_c$ be the number of classes in a project for which \textsc{Sensor} could construct instances; $N_t$ be the total number of classes in this project; and $N_i$ be the average number of instances with different divergence arguments constructed for each class in a project. 
%\civi{I don't understand the experimental setting of this part. Will Sensor try to construct instances for each class in a project?}
Then, $N_c/N_t$ describes \textsc{Sensor}'s capability on constructing class instances with extracted invocation contexts.
In our ground truth dataset, the 225 isomerous conflicting API pairs are selected from 123 Java projects.
As shown in Figures~\ref{fig_122}(a) and~\ref{fig_122}(b), the box plots show the distribution of indicators $N_c/N_t$ and $N_i$ in these projects.
On average, \textsc{Sensor} could construct instances with extracted arguments for 76.8\% of classes in the projects.
We looked into the code and found that \textsc{Sensor} could not instantiate the remaining classes mainly because their required arguments are not provided in the source code, or for the accessibility reason.
On average, \textsc{Sensor} constructed 3.13 instances for each class with divergence arguments, in these projects.
Specially, in project {\mycode FluentLenium-3.9.0}, there are 89 classes having more than ten invocation contexts for their corresponding constructors.
The diverse constructed class instances significantly increase the probability of capturing inconsistent behaviors with different invocation contexts.

Let $Rp$ represent the number of class instances that are successfully seeded by \textsc{Sensor} in a test case, and $N_o$ be the total number of instantiated classes in this test case.
Then, $Rp/ N_o$ is the substitution rate of class instances by our approach in a test case. 
Figures~\ref{fig_122}(c) and~\ref{fig_122}(d) show the values of $Rp/ N_o$ in the 596 generated test cases that successfully captured the inconsistent behaviors and the 264 test cases that caused crashes before triggering the conflicting APIs, over the ten runs, respectively.
The average value of $Rp/ N_o$ in Figure~\ref{fig_122}(c) is 0.22 higher than that in Figure~\ref{fig_122}(d).
%\civi{I don't think the results are convincing since the median value in figure c is lower than that in figure d.}
The results demonstrated the validity of our seeded class instances.
Furthermore, suppose that $Argu_s$ is the number of arguments required by a constructor that can be extracted from the source code and $Argu_t$ is the total number of required arguments in this constructor.
%Then, $Argu_s/ Argu_t$ indicates the correctness of the state of a constructed class instance.
For the seeded class instances in the 596 test cases that successfully identified SC issues, the average value of $Argu_s/ Argu_t$ is 0.24 higher than that of class instances in the 264 test cases causing program crashes. %\civi{provide the value of the failing test cases.}
This validates the correctness of the arguments extracted by \textsc{Sensor} for constructing class instances.
%\end{comment}

%\vspace{-2mm}
\subsubsection{Effectiveness on industrial projects.}

To further assess \textsc{Sensor}'s effectiveness, we applied it to the industrial projects in the {\mycode Neusoft} Co. Ltd (SSE: 600718) and received an assessment report~\cite{Assessment}.
{\mycode Neusoft}~\cite{Neusoft} is the largest IT solutions \& services provider in China, which has considerable large-scale Java projects with hundreds of third party libraries.
Diagnosing SC issues is one of the key challenges for their developers.
%We applied \textsc{Sensor} to ten Java projects of the \emph{XIKANG} department in \emph{Neusoft}.
%Ideally, the original developers' domain knowledge of the projects are the best criterion for evaluating the precision of SC issue detection.
%To achieve this, we interviewed 9 original developers of the corresponding projects, and asked for their verification for the detected SC issues.
Table~3 reports the results of applying \textsc{Sensor} to ten industrial subjects that comprise over 0.58 million lines of code.
We invited nine developers who participated in development of the selected projects to verify the detected SC issues. %\scc{Why 9 developers for 10 projects?} %, under the assumption that the developers' domain knowledge of the projects is the ideal criterion to validate those detected SC issues. %is the ideal criterion for evaluating the precision of SC issue detection.
We did not evaluate the $Recall$ in this experiment since it is difficult to obtain the complete set of SC issues in the projects.

In Table~3, columns ``$CA$'' and ``$ICA$'' represent the number of conflicting API pairs and isomerous conflicting API pairs caused by two conflicting library versions in the project, respectively.
%From the results, we found that $ICA$ is only a small percentage of $CA$ and the ratio of $ICA$ to $CA$ is affected by the release interval between the two conflicting library versions in general. %\civi{I cannot follow the later part of this sentence.}
%For instance, in project {\mycode P1}, ASTs are inconsistent among 42.7\% of the conflicting API pairs, as the interval between the introduced conflicting library version is large (\ie containing 13 different releases).
%As a result, this issue aroused developers' great attention to solve it.
Among the 56 detected SC issues, 46 were confirmed by developers as true positives (TP) and 10 were labeled as false positives, leading to a $Precision$ of 0.821. % on the industrial projects.
By communicating with the projects' developers,
we found that the main cause of the above false positives (FP) is the same as that in the open-source projects.
In these cases, the inconsistent variable states affected by the conflicting API pairs are benign for the semantics of industrial subjects.
In particular, we received positive feedback from the {\mycode Neusoft}'s testing team, on the high precision of \textsc{Sensor}.
Such results indicate that \textsc{Sensor} not only achieves significant effectiveness on open-source projects, but also performs great on industrial subjects.

\begin{table}[t!]
	\label{Tab3}
	%\footnotesize
	\scriptsize
	%	\vspace{-4mm}
	\caption{The results on anonymized industry projects}
	\bgroup
	\setlength\tabcolsep{12.5pt}     % horizental space
	\def\arraystretch{1.1}
	%	\vspace{-4mm}
	\begin{tabular}{c|c|c|c|c|c}
		\toprule
		%	\hline
		Projects                  & SLOC                                & $CA$  & $ICA$ & $TP$ & $FP$ \\ \hline
		P1                  & \textgreater{}100K                  & 103 & 44  & 7  & 2  \\ \hline
		P2                  & \textgreater{}100K                  & 78  & 9   & 3  & 0  \\ \hline
		P3                  & \textgreater{}100K                  & 213 & 36  & 10 & 2  \\ \hline
		\multirow{2}{*}{P4} & \multirow{2}{*}{\textgreater{}100K} & 52  & 20  & 5  & 2  \\ \cline{3-6} 
		&                                     & 89  & 25  & 6  & 1  \\ \hline
		P5                  & \textgreater{}50K                   & 31  & 7   & 2  & 1  \\ \hline
		\multirow{2}{*}{P6} & \multirow{2}{*}{\textgreater{}50K}  & 64  & 12  & 3  & 0  \\ \cline{3-6} 
		&                                     & 45  & 11  & 3  & 1  \\ \hline
		P7                  & \textgreater{}20K                   & 23  & 7   & 2  & 0  \\ \hline
		P8                  & \textgreater{}20K                   & 42  & 8   & 3  & 0  \\ \hline
		P9                  & \textgreater{}20K                   & 15  & 4   & 1  & 1  \\ \hline
		P10                 & \textgreater{}20K                   & 7   & 2   & 1  & 0  \\ \hline
		$Precision$           & \multicolumn{5}{c}{46 / 56 = 0.821}                        \\  \bottomrule%\hline
	\end{tabular}
	\egroup
	%\vspace{-4mm}
\end{table}

%\begin{comment}
\begin{table}[t!]
	\label{Tab4}
	%	\footnotesize
	%	\vspace{-2mm}
	\caption{The SC issues reported by \textsc{Sensor}}
	\bgroup
	\setlength\tabcolsep{0.002pt}     % horizental space
	\def\arraystretch{1.2}
	%\def\arraystretch{0.89}
	%	\vspace{-4mm}
	%\definecolor{Status2}{HTML}{5ab4ac}
	%\definecolor{Status3}{HTML}{c7eae5}
	%\definecolor{Status4}{HTML}{f5f5f5}
	\setlength\fboxsep{1pt}
	\scriptsize
	\begin{tabular}{|l|}
		\hline
		\begin{tabular}[c]{@{}l@{}}{Htm.java, \#550, 1}; {EasyTransaction, \#144, 1}; {Hydra, \#364$\spadesuit$, 1}; {Motan, \#800$\bigstar$, 15}; \\{Restx, \#297$\bigstar$, 3}; {Netty-rest, \#8$\bigstar$, 12}; {Netty-rest, \#9$\bigstar$, 9}; {Aws-sdk-java, \#1897$\lozenge$, 1}; \\{Ff4j,\#336$\bigstar$, 10}; {Retrofit, \#3018$\bigstar$, 12}; {Guagua,\#103$\bigstar$, 19}; {Wechat-springmvc, \#17$\bigstar$, 3} \\{Jss7, \#309, 1};
			{Motan, \#809$\bigstar$, 15}; {Product-iots, \#1911, 1}; {Nutzboot, \#199$\bigstar$, 2}; \\{Atom-hopper, \#301$\bigstar$, 1}; {Quick-media, \#41$\bigstar$, 1}; {Ontop, \#287$\bigstar$, 12}; {Ontop, \#288$\bigstar$, 17}; \\{Odo, \#173$\bigstar$, 1};
			{Openstack-java-sdk, \#214$\bigstar$, 1};  {Java-design-patterns, \#868$\bigstar$, 1}; \\{Hmily, \#86$\bigstar$, 1}; {Ninja, \#654$\bigstar$, 1}; {Javacpp, \#295$\lozenge$, 1}; {FastjsonExploit, \#6$\bigstar$, 1}; \\{MiA,\#11$\spadesuit$, 1}; {Vertx-examples, \#335$\spadesuit$, 1}; {Vertx-examples, \#336$\spadesuit$, 1}; {Yawp, \#121$\bigstar$, 1};  \\ {Apache/Hive, \#21374$\bigstar$, 1}; {Rest-assured, \#1143$\bigstar$, 1} \end{tabular} \\ \hline
		\multicolumn{1}{l}{$\langle$ Project name, issue report ID, the number of SC issues in the issue report $\rangle$}\\
		\multicolumn{1}{l}{$\bigstar$: The issues have already been fixed. $\spadesuit$: The issues were confirmed and in}\\
		\multicolumn{1}{l}{the process of being fixed. $\lozenge$: False positive cases.}\\ 
		\multicolumn{1}{l}{The detailed information is available at: https://sensordc.github.io/}\\[-0.03cm] 
		%\multicolumn{1}{l}{} \\[-0.02cm] 
	\end{tabular}
	\egroup
	%\vspace{-1mm}
\end{table}

\vspace{-3mm}
\subsection{RQ2: Usefulness of Sensor}
\label{sec:Usefulness}

\textsc{Sensor} successfully detected 150 SC issues from 29 projects among all the 92 projects.
Note that the SC issues caused by a pair of conflicting library versions were merged into one issue report.
Altogether, we submitted 33 issue reports.
As shown in Table~4, 27 out of 33 issue reports (81.8\%) were confirmed by developers as real bugs;
23 out of 27 confirmed reports (85.2\%) were quickly fixed; and 4 of them (14.8\%) were in the process of being fixed.
Among the seven unconfirmed issue reports, two were labeled as false positives and the others are not confirmed mainly due to inactive maintenance of the corresponding release versions.

%The reasons of the false positive causes, are similar as those we previously discussed.
%The developers considered the reported issues as benign after checking the impacts of our reported semantic differences.
%For example in {\mycode Issue \#1897}~\cite{Issue1897}, the isomerous conflicting API pair $<${\mycode hashCode():Junit.4.4}, {\mycode hashCode():Junit.4.12}$>$ indirectly affects the state of an {\mycode HashMap} variable used in an entry method of project {\mycode Aws-sdk-java}.
%Methods {\mycode hashCode():Junit.4.4} and {\mycode hashCode():Junit.4.12} differ in the way of constructing the value {\mycode Key}, but both versions of this method can ensure the uniqueness of {\mycode Key} used to map its associated {\mycode Value}.
%So, the developer considered it a false positive as it would not introduce bugs.

%\vspace{-2mm}
%\subsubsection{Developers' feedback from open-source community.}

We have received developers' positive feedbacks on the reported SC issues and our tool.
Developers in {\mycode Issue \#11}~\cite{Issue11} agreed that the provided test case indeed triggered realistic program behaviors, which has  facilitated their diagnosis for semantic conflicts.
In particular, a developer confirmed the usefulness of our approach :

\vspace{-0.5mm}
\emph{``I encountered the same problem when using {\mycode MiA}.
	I just noticed something strange happened in the [min, max] range of operation progresses.
	Thanks for your test case. It helped me reproduce this issue.
	By amazing coincidence, I got the similar outputs as your test.''}

Besides, developers have expressed great interests in our detection technique for SC issues.
For instance, in {\mycode Issue \#288}~\cite{Issue288}, an experienced developer~\cite{Developer} in the {\mycode Ontop} community has been looking for a technique of such kind to detect semantic conflicts:
%\textsc{Sensor} attracts his attention, so that we received the following feedback:

%\vspace{-0.9mm}

%\begin{comment}
\emph{``
	I am very interested in the detection method so that in the future we will have a more systematic approach to avoid the issue related to such conflict, which is in general quite subtle and difficult to debug.''}
%\end{comment}

%\vspace{-2mm}
%\subsubsection{Developers' feedback from industry.}
\textsc{Sensor} was highly recognized in the assessment report~\cite{Assessment} provided by {\mycode Neusoft}:

\emph{``On average, it took about 20.5 hours to obtain the diagnosis report for a large-scale {\mycode Neusoft} project, and the run time depends on the number of conflicting APIs in the project.
	Although the testing task is time-consuming, \textsc{Sensor} did a great job in automatically detecting SC issues.
	The generated diagnosis reports indeed helped us identify many issues that could hardly be found using our existing test suites.''}

The above results and developers' feedback demonstrate that the information (e.g., test cases) provided by \textsc{Sensor} is useful for developers to diagnose the SC issues in practice.

%\vspace{-3mm}

\vspace{-3mm}
\subsection{Discussions}
\label{sec:Discussion}

We further analyzed the root causes and distributions of the behavioral inconsistencies induced by the identified isomerous conflicting API pairs of all the subjects as shown in Table~4.
Statistically, for the 150 isomerous conflicting API pairs that affect the client projects' program behaviors, we further categorized the exposed behavioral inconsistencies into three types:
(a) 135 of them (90.0\%) only cause \emph{variable state inconsistencies};
(b) 5 of them (3.3\%) only lead to \emph{test outcome inconsistencies}; (c) 10 of them (6.7\%) result in both \emph{variable state} and \emph{test outcome} inconsistencies.

%\begin{figure}[H]
%	\vspace{-2mm}
%	\centering
%	\includegraphics[width=0.4\textwidth]{./figure/F_14.pdf}
%	\vspace{-2mm}
%	\caption{Distributions of the behavioral inconsistencies}	
%	\label{fig_q}
%	\vspace{-3mm}
%\end{figure}

%\vspace{-0mm}
%\begin{figure*}[h]
%	\centering
%	\includegraphics[width=0.96\textwidth]{./figure/771.pdf}
%	\vspace{-1mm}
%	\caption{Examples of inconsistent program behaviors}	
%	\label{fig_e}
%	\vspace{-4mm}
%\end{figure*}

By manually examining the source code of isomerous conflicting API pairs, we found that \emph{variable state inconsistencies} are mainly caused by adding or deleting control branches in one version of the conflicting API (\eg issue \#550~\cite{Issue550}) or inconsistent function implementations between conflicting API pairs (\eg issue \#1143~\cite{Issue1143}).
In addition, strengthening the precondition or weakening postcondition of a referenced method will lead to \emph{test outcome inconsistencies}.
A method's precondition is the condition that a caller must satisfy before calling the method, and a method's postcondition is the condition that a callee must satisfy before returning from the method~\cite{jezek2015java}.
For instance, in issue \#9~\cite{Issue9}, replacing the shadowed version of a method with the loaded version resulted in strengthening the precondition of this conflicting method. 
Therefore, the caller in the host project will trigger a {\mycode NullPointerException} when the caller of method {\mycode setNamesSize()} in the client project passes an empty {\mycode HashSet} to it.
Similarly, in issue \#809~\cite{Issue809}, referencing the actually loaded version of a method to substitute the shadowed version will weaken its postcondition, which can trigger a crash in the caller.
The above cases will break the compatibility of libraries in the client projects.

\vspace{-2mm}
\section{Threats to Validity}
\label{sec:Threats To Validity}
\noindent\textbf{Ground truth dataset collection.} 
Collecting the ground truth dataset of SC issues is challenging and can be a threat to the evaluation results.
To avoid introducing noises in our dataset, we upgraded/ downgraded the actually-loaded library versions in a series of Java projects.
After altering the library versions, we selected the conflicting API pairs that could trigger the {\mycode AssertionErrors} when executing the projects' associated tests, as the cases that definitely caused SC issues.
%In addition, we mined a collection of semantic-preserving library version changes, which did not cause any semantic issues in the next 24 months after their corresponding commits are merged.
%We then labeled the isomerous conflicting API pairs that can be triggered by a continuous integration build tool without errors when submitting the above commits, as the ones would not introduce SC issues into their client projects.

\noindent\textbf{Validity of developers' feedback.} In this paper, we rely on developers' feedback to validate the effectiveness and usefulness of \textsc{Sensor} on both industrial and open source projects.
However, there might be different opinions towards the validity of the issue reports for different developers.
To mitigate such threat, for the industrial subjects, we invited nine original developers with the domain knowledge of the selected projects for verification.
%under the assumption that the developers' domain knowledge of the projects is the ideal criterion for evaluation the precision of SC issue detection.
For the open source projects, we did not encounter the controversies for all the evaluated subjects. 
Therefore, the received feedback demonstrate the effectiveness and usefulness of our approach.

\vspace{-1mm}
\section{Related Work}
\label{sec:Related work}
%\vspace{-2mm}
\noindent\textbf{Dependency conflict.}
Library conflicts are challenging to detect for a program analysis and difficult to avoid for library developers.
Determining whether two or more libraries cannot be built together is an important issue in the quality assurance process of software projects.
Blincoe et at.~\cite{dietrich2019dependency} conducted an in-depth study of millions of dependencies across multiple software ecosystems.
They found that using a range of versions to declare dependencies could facilitate the automated repairing for dependency conflict issues, when adopting semantic versioning strategies.
Yet, since the vast majority of Java projects declare the fixed versions of their referenced third party libraries, semantic versioning does not play a major role in repairing dependency conflict issues in the Java ecosystem.

Ghorbani et al.~\cite{ghorbani2019detection} formally defined eight inconsistent modular dependencies that may arise in Java-9 applications, and proposed a technique \textsc{Darcy} to detect and repair such specified inconsistent dependencies.
So far, there are a significant fraction of Java projects that do not adopt Java-9 mechanism.
Therefore, an effective approach to diagnosing the SC issues is still urgently needed in the Java ecosystem.

Suzaki et al.' approach~\cite{artho2012software} mainly focused on conflicts on resource access, conflicts on configuration data, and interactions between uncommon combinations of packages and categorized them to provide useful suggestions on how to prevent and detect such problem.
%In their approach, they 
Patra et al.~\cite{patra2018conflictjs} were the first researchers studying the detection strategy for conflicts among JavaScript libraries.
They tackled the huge search space of possible conflicts in two phases, i.e., identifying potentially conflicting library pairs and synthesizing library clients to validate conflicts.
Soto-Valero et al.~\cite{soto2019emergence} presented quantitative empirical evidence about how the immutability of artifacts in {\mycode Maven Central} supports the emergence of natural software diversity.
%They analyzed 1,487,956 artifacts and observed that more than 30\% of libraries have multiple versions that are actively used by latest artifacts.
Wang et al.~\cite{wang2018dependency} conducted a study to characterize the manifestations of dependency conflicts in Java projects, and presented an automated technique to diagnose dependency conflict issues.
Afterwards, they developed \textsc{Riddle} to generate tests to collect crashing stack traces to facilitate dependency conflict diagnosis~\cite{WANG2019STACK}.
However, there is no previous work analyzing the impacts on semantic behaviors of programs when dependency conflicts happen.

\noindent\textbf{Differential testing and analysis.}
%Many automated testing techniques~\cite{rojas2016seeding, fraser2012seed, bell2018d, rosner2014bounded, alipour2016generating, ma2015grt, soltana2017synthetic, christakis2015ic, misurda2005demand, rocha2015memory, zhang2012compositional, wang2018adversarial, abad2013improving, fraser2012mutation, conroy2007automatic, zhang2016isomorphic, arcuri2017private, robinson2014automatic, gligoric2010test, ma2018deepmutation, hao2010test, staats2012danger, nguyen2014guitar, xiao2013characteristic, andrews2007nighthawk, chen2018coverage} have been proposed to detect software bugs, whereas it is difficult to define the oracle of generated tests without prior knowledge of expected outputs~\cite{columbus}.
%Differential testing addresses this problem by examining test outcomes of comparable systems.
%Differential testing uses similar programs as crossreferencing oracles to find semantic bugs that do not exhibit
%explicit erroneous behaviors like crashes or assertion failures.
Differential testing and analysis techniques have been used to find bugs across many types of programs~\cite{zhang2017automated, chapman2011automated, chen2016coverage, yang2011finding, chowdhury2016cyfuzz, daniel2007automated, groce2007randomized, lammel2006controllable, cadar2008klee, ray2013detecting, zhong2013exposing, lin2014detecting}.
Zhang et al. \cite{zhang2016isomorphic} implemented an isomorphic regression testing approach named \textsc{Ison}, which compares the behaviors of modified programs to check whether abnormal behaviors are induced in the new code versions. 
Xie et al.~\cite{xie2007towards} presented a differential unit testing technique, \textsc{Diffut}, which compared the methods between different program revisions.
Petsios et al. \cite{petsios2017nezha} proposed an effective technique \textsc{Nezha} to trigger semantic bugs, using gray-box and black-box mechanisms to generate inputs for differential testing.
\textsc{Nezha} is applicable to detecting the semantic differences among different libraries providing similar functionalities.
Compared with the above differential testing techniques, \textsc{Sensor} has a different goal: detecting semantic conflicts combining host projects' invocation contexts.

\noindent\textbf{Test input generation.} Existing automated testing generation approaches use many techniques to create inputs for exercising a software under test with minimal human efforts, including feedback-directed random test generation~\cite{artzi2011framework, pacheco2007feedback, pradel2012fully}, search-based techniques~\cite{chiang2014efficient, fraser2011evosuite}, seeding strategies~\cite{alshahwan2011automated, rojas2016seeding, fraser2012seed, sakti2015instance, mcminn2012search, alshraideh2006search}, and symbolic reasoning-based test generators~\cite{cadar2008klee, godefroid2005dart, sen2005cute, thummalapenta2011synthesizing}.
Xu et al.~\cite{xu2013mining} presented a mining approach to building a decision tree model according to the test inputs generated from Java bytecode.
It converts Java bytecode into the Jimple representation, extracts predicates from the control flow graph of the Jimple code, and uses these predicates as attributes for organizing training data to build a decision tree.
Dallmeier et al.~\cite{dallmeier2010generating} proposed an improved dynamic specification mining technique, \textsc{TAUTOKO}, to generate test cases.
Since previous specification mining technique entirely depends on the observed executions, the resulting specification may be too incomplete to be useful if not enough tests are available.
To address this problem, \textsc{TAUTOKO} explores previously unobserved aspects of the execution space.
Their evaluation results shown that the enriched specifications cover more general behaviors and much more exceptional behaviors.
Toffola et al.~\cite{della2017saying} proposed an approach to extract literals from thousands of tests and to adapt information retrieval techniques to find values suitable for a particular domain. 

Despite all successes, test generation still suffers from non-trivial limitations in exposing SC issues.
First, approach~\cite{xu2013mining} are more effective to deal with the code snippets with simple data types that can be easily convert into Jimple representations.
While our empirical study results show that to trigger the real SC issues, effective test generation techniques should have the ability to construct divergence arguments for parameterized complex constructors.
Second, the effectiveness of approaches~\cite{dallmeier2010generating, della2017saying} entirely depends on the detection ability of existing test suites of projects under test.
Our empirical study provides evidences that combining invocation contexts of constructors in source code can effectively improve the possibility of capturing inconsistent program behaviors caused by SC issues.
In addition, existing techniques~\cite{alshahwan2011automated, rojas2016seeding, fraser2012seed, sakti2015instance, mcminn2012search, alshraideh2006search} have proposed different seeding strategies for test input generation, especially for strings and primitive types.
Since these strategies generate constructor arguments without considering their invocation contexts, they are not effective in constructing valid class instances to expose conflicting API pairs.
\textsc{Sensor} adopts a new seeding strategy of class instances inspired by the our empirical findings summarized in Section~\ref{sec:Empirical}.
\textsc{Sensor} injects the invocation context information extracted from the source code into class instances with the aim of generating divergence arguments.

%\textbf{Automated test generation.}

\vspace{-1mm}
\section{Conclusion and Future Work}
In this paper, we presented an effective and automated test generation technique \textsc{Sensor}, 
which are capable of producing valid inputs to trigger the SC issues.
The evaluation results show that \textsc{Sensor} can achieve a $Precision$ of 0.803 and a $Recall$ of 0.760 on open source projects and a $Precision$ of 0.821 on industrial subjects.
\textsc{Sensor} has detected 150 SC issues from 29 open source projects and submitted 33 issue reports to them.
Encouragingly, 27 issue reports (81.8\%) have been confirmed by developers as real SC issues.
% with the confirmation of developers.
Although \textsc{Sensor} is designed for detecting SC issues, it can be adapted to other problems arising from library evolution (e.g., safeguarding the reliability upgrading libraries). 
In future, we plan to combine symbolic execution or fuzzing techniques with our technique to improve its test input exploration capability.
%\wu{Shorten this section. We do not need to emphasize the evaluation in this section. }

% or


\ifCLASSOPTIONcaptionsoff
  \newpage
\fi



%\bibliographystyle{IEEEtran}

\bibliography{bibliography}{}

\begin{thebibliography}{10}
\providecommand{\url}[1]{#1}
\csname url@samestyle\endcsname
\providecommand{\newblock}{\relax}
\providecommand{\bibinfo}[2]{#2}
\providecommand{\BIBentrySTDinterwordspacing}{\spaceskip=0pt\relax}
\providecommand{\BIBentryALTinterwordstretchfactor}{4}
\providecommand{\BIBentryALTinterwordspacing}{\spaceskip=\fontdimen2\font plus
\BIBentryALTinterwordstretchfactor\fontdimen3\font minus
  \fontdimen4\font\relax}
\providecommand{\BIBforeignlanguage}[2]{{%
\expandafter\ifx\csname l@#1\endcsname\relax
\typeout{** WARNING: IEEEtran.bst: No hyphenation pattern has been}%
\typeout{** loaded for the language `#1'. Using the pattern for}%
\typeout{** the default language instead.}%
\else
\language=\csname l@#1\endcsname
\fi
#2}}
\providecommand{\BIBdecl}{\relax}
\BIBdecl

\bibitem{bao2018inference}
L.~Bao, Z.~Xing, X.~Xia, D.~Lo, and A.~E. Hassan, ``Inference of development
  activities from interaction with uninstrumented applications,''
  \emph{Empirical Software Engineering}, vol.~23, no.~3, pp. 1313--1351, 2018.

\bibitem{jezek2015java}
K.~Jezek, J.~Dietrich, and P.~Brada, ``How java apis break--an empirical
  study,'' \emph{Information and Software Technology}, vol.~65, pp. 129--146,
  2015.

\bibitem{teyton2014study}
C.~Teyton, J.-R. Falleri, M.~Palyart, and X.~Blanc, ``A study of library
  migrations in java,'' \emph{Journal of Software: Evolution and Process},
  vol.~26, no.~11, pp. 1030--1052, 2014.

\bibitem{macho2018automatically}
C.~Macho, S.~McIntosh, and M.~Pinzger, ``Automatically repairing
  dependency-related build breakage,'' in \emph{2018 IEEE 25th International
  Conference on Software Analysis, Evolution and Reengineering (SANER)}.\hskip
  1em plus 0.5em minus 0.4em\relax IEEE, 2018, pp. 106--117.

\bibitem{wang2018dependency}
Y.~Wang, M.~Wen, Z.~Liu, R.~Wu, R.~Wang, B.~Yang, H.~Yu, Z.~Zhu, and S.-C.
  Cheung, ``Do the dependency conflicts in my project matter?'' in
  \emph{Proceedings of the 2018 26th ACM Joint Meeting on European Software
  Engineering Conference and Symposium on the Foundations of Software
  Engineering}.\hskip 1em plus 0.5em minus 0.4em\relax ACM, 2018, pp. 319--330.

\bibitem{liang1998dynamic}
S.~Liang and G.~Bracha, ``Dynamic class loading in the java [tm] virtual
  machine,'' \emph{Acm sigplan notices}, vol.~33, pp. 36--44, 1998.

\bibitem{WANG2019STACK}
Y.~Wang, M.~Wen, R.~Wu, Z.~Liu, S.~H. Tan, Z.~Zhu, H.~Yu, and S.-C. Cheung,
  ``{Could I Have a Stack Trace to Examine the Dependency Conflict Issue?}'' in
  \emph{{Proceedings of the 41th International Conference on Software
  Engineering}}, ser. ICSE.\hskip 1em plus 0.5em minus 0.4em\relax ACM/IEEE,
  2019.

\bibitem{wang2018adversarial}
J.~Wang, G.~Dong, J.~Sun, X.~Wang, and P.~Zhang, ``Adversarial sample detection
  for deep neural network through model mutation testing,'' \emph{arXiv
  preprint arXiv:1812.05793}, 2018.

\bibitem{Issue214}
``Issue \#214 of project openstack-java-sdk,''
  \url{https://github.com/woorea/openstack-java-sdk/issues/214}, 2020,
  accessed: 2020-01-31.

\bibitem{PRIssue214}
``A pr of issue \#214 in project openstack-java-sdk,''
  \url{https://github.com/woorea/openstack-java-sdk/pull/215}, 2020, accessed:
  2020-01-31.

\bibitem{Neusoft}
``Neusoft,'' \url{https://www.neusoft.com/}, 2020, accessed: 2020-01-31.

\bibitem{Rest-assured}
``Rest-assured,'' \url{https://github.com/rest-assured/rest-assured}, 2020,
  accessed: 2020-01-31.

\bibitem{Java-design-patterns}
``Java-design-patterns,''
  \url{https://github.com/iluwatar/java-design-patterns}, 2020, accessed:
  2020-01-31.

\bibitem{Hmily}
``Hmily,'' \url{https://github.com/yu199195/hmily}, 2020, accessed: 2020-01-31.

\bibitem{artzi2008recrash}
S.~Artzi, S.~Kim, and M.~D. Ernst, ``Recrash: Making software failures
  reproducible by preserving object states,'' in \emph{European conference on
  object-oriented programming}.\hskip 1em plus 0.5em minus 0.4em\relax
  Springer, 2008, pp. 542--565.

\bibitem{falleri2014fine}
J.-R. Falleri, F.~Morandat, X.~Blanc, M.~Martinez, and M.~Monperrus,
  ``Fine-grained and accurate source code differencing,'' in \emph{Proceedings
  of the 29th ACM/IEEE international conference on Automated software
  engineering}.\hskip 1em plus 0.5em minus 0.4em\relax ACM, 2014, pp. 313--324.

\bibitem{schroter2010stack}
A.~Schroter, A.~Schr{\"o}ter, N.~Bettenburg, and R.~Premraj, ``Do stack traces
  help developers fix bugs?'' in \emph{2010 7th IEEE Working Conference on
  Mining Software Repositories (MSR 2010)}.\hskip 1em plus 0.5em minus
  0.4em\relax IEEE, 2010, pp. 118--121.

\bibitem{sakti2015instance}
A.~Sakti, G.~Pesant, and Y.-G. Gu{\'e}h{\'e}neuc, ``Instance generator and
  problem representation to improve object oriented code coverage,'' \emph{IEEE
  Transactions on Software Engineering}, vol.~41, no.~3, pp. 294--313, 2015.

\bibitem{fraser2016evosuite}
G.~Fraser and A.~Arcuri, ``Evosuite at the sbst 2016 tool competition,'' in
  \emph{Proceedings of the 9th International Workshop on Search-Based Software
  Testing}, 2016, pp. 33--36.

\bibitem{arcuri2014automated}
A.~Arcuri, G.~Fraser, and J.~P. Galeotti, ``Automated unit test generation for
  classes with environment dependencies,'' in \emph{Proceedings of the 29th
  ACM/IEEE international conference on Automated software engineering}, 2014,
  pp. 79--90.

\bibitem{kim2006long}
S.~Kim and E.~J. Whitehead~Jr, ``How long did it take to fix bugs?'' in
  \emph{Proceedings of the 2006 international workshop on Mining software
  repositories}.\hskip 1em plus 0.5em minus 0.4em\relax ACM, 2006, pp.
  173--174.

\bibitem{wu2011relink}
R.~Wu, H.~Zhang, S.~Kim, and S.-C. Cheung, ``Relink: recovering links between
  bugs and changes,'' in \emph{Proceedings of the 19th ACM SIGSOFT symposium
  and the 13th European conference on Foundations of software
  engineering}.\hskip 1em plus 0.5em minus 0.4em\relax ACM, 2011, pp. 15--25.

\bibitem{Assessment}
``Assessment report,'' \url{https://sensordc.github.io/}, 2020, accessed:
  2020-01-31.

\bibitem{Issue11}
``Issue \#11 of project mia,'' \url{https://github.com/tdunning/MiA/issues/11},
  2020, accessed: 2020-01-31.

\bibitem{Issue288}
``Issue \#288 of project ontop,''
  \url{https://github.com/ontop/ontop/issues/288}, 2020, accessed: 2020-01-31.

\bibitem{Developer}
``An experienced developer of project ontop,'' \url{https://github.com/ghxiao},
  2020, accessed: 2020-01-31.

\bibitem{Issue550}
``Issue \#550 of project htm.java,''
  \url{https://github.com/numenta/htm.java/issues/550}, 2020, accessed:
  2020-01-31.

\bibitem{Issue1143}
``Issue \#1143 of project rest-assured,''
  \url{https://github.com/rest-assured/rest-assured/issues/1143}, 2020,
  accessed: 2020-01-31.

\bibitem{Issue9}
``Issue \#9 of project netty-rest,''
  \url{https://github.com/buremba/netty-rest/issues/9}, 2020, accessed:
  2020-01-31.

\bibitem{Issue809}
``Issue \#809 of motan,'' \url{https://github.com/weibocom/motan/issues/809},
  2020, accessed: 2020-01-31.

\bibitem{dietrich2019dependency}
J.~Dietrich, D.~Pearce, J.~Stringer, A.~Tahir, and K.~Blincoe, ``Dependency
  versioning in the wild,'' in \emph{2019 IEEE/ACM 16th International
  Conference on Mining Software Repositories (MSR)}.\hskip 1em plus 0.5em minus
  0.4em\relax IEEE, 2019, pp. 349--359.

\bibitem{ghorbani2019detection}
N.~Ghorbani, J.~Garcia, and S.~Malek, ``Detection and repair of architectural
  inconsistencies in java,'' in \emph{2019 IEEE/ACM 41st International
  Conference on Software Engineering (ICSE)}.\hskip 1em plus 0.5em minus
  0.4em\relax IEEE, 2019, pp. 560--571.

\bibitem{artho2012software}
C.~Artho, K.~Suzaki, R.~Di~Cosmo, R.~Treinen, and S.~Zacchiroli, ``Why do
  software packages conflict?'' in \emph{Proceedings of the 9th IEEE Working
  Conference on Mining Software Repositories}.\hskip 1em plus 0.5em minus
  0.4em\relax IEEE Press, 2012, pp. 141--150.

\bibitem{patra2018conflictjs}
J.~Patra, P.~N. Dixit, and M.~Pradel, ``Conflictjs: Finding and understanding
  conflicts between javascript libraries,'' in \emph{Proceedings of the 40th
  International Conference on Software Engineering}.\hskip 1em plus 0.5em minus
  0.4em\relax ACM, 2018, pp. 741--751.

\bibitem{soto2019emergence}
C.~Soto-Valero, A.~Benelallam, N.~Harrand, O.~Barais, and B.~Baudry, ``The
  emergence of software diversity in maven central,'' in \emph{MSR 2019-16th
  International Conference on Mining Software Repositories}.\hskip 1em plus
  0.5em minus 0.4em\relax ACM, 2019, pp. 1--11.

\bibitem{zhang2017automated}
T.~Zhang and M.~Kim, ``Automated transplantation and differential testing for
  clones,'' in \emph{Proceedings of the 39th International Conference on
  Software Engineering}.\hskip 1em plus 0.5em minus 0.4em\relax IEEE, 2017, pp.
  665--676.

\bibitem{chapman2011automated}
P.~Chapman and D.~Evans, ``Automated black-box detection of side-channel
  vulnerabilities in web applications,'' in \emph{Proceedings of the 18th ACM
  conference on Computer and communications security}.\hskip 1em plus 0.5em
  minus 0.4em\relax ACM, 2011, pp. 263--274.

\bibitem{chen2016coverage}
Y.~Chen, T.~Su, C.~Sun, Z.~Su, and J.~Zhao, ``Coverage-directed differential
  testing of jvm implementations,'' in \emph{ACM SIGPLAN Notices}, vol.~51,
  no.~6.\hskip 1em plus 0.5em minus 0.4em\relax ACM, 2016, pp. 85--99.

\bibitem{yang2011finding}
X.~Yang, Y.~Chen, E.~Eide, and J.~Regehr, ``Finding and understanding bugs in c
  compilers,'' in \emph{ACM SIGPLAN Notices}, vol.~46, no.~6.\hskip 1em plus
  0.5em minus 0.4em\relax ACM, 2011, pp. 283--294.

\bibitem{chowdhury2016cyfuzz}
S.~A. Chowdhury, T.~T. Johnson, and C.~Csallner, ``Cyfuzz: A differential
  testing framework for cyber-physical systems development environments,'' in
  \emph{International Workshop on Design, Modeling, and Evaluation of Cyber
  Physical Systems}.\hskip 1em plus 0.5em minus 0.4em\relax Springer, 2016, pp.
  46--60.

\bibitem{daniel2007automated}
B.~Daniel, D.~Dig, K.~Garcia, and D.~Marinov, ``Automated testing of
  refactoring engines,'' in \emph{Proceedings of the the 6th joint meeting of
  the European software engineering conference and the ACM SIGSOFT symposium on
  The foundations of software engineering}.\hskip 1em plus 0.5em minus
  0.4em\relax ACM, 2007, pp. 185--194.

\bibitem{groce2007randomized}
A.~Groce, G.~Holzmann, and R.~Joshi, ``Randomized differential testing as a
  prelude to formal verification,'' in \emph{29th International Conference on
  Software Engineering (ICSE'07)}.\hskip 1em plus 0.5em minus 0.4em\relax IEEE,
  2007, pp. 621--631.

\bibitem{lammel2006controllable}
R.~L{\"a}mmel and W.~Schulte, ``Controllable combinatorial coverage in
  grammar-based testing,'' in \emph{IFIP International Conference on Testing of
  Communicating Systems}.\hskip 1em plus 0.5em minus 0.4em\relax Springer,
  2006, pp. 19--38.

\bibitem{cadar2008klee}
C.~Cadar, D.~Dunbar, D.~R. Engler \emph{et~al.}, ``Klee: Unassisted and
  automatic generation of high-coverage tests for complex systems programs.''
  in \emph{OSDI}, vol.~8, 2008, pp. 209--224.

\bibitem{ray2013detecting}
B.~Ray, M.~Kim, S.~Person, and N.~Rungta, ``Detecting and characterizing
  semantic inconsistencies in ported code,'' in \emph{2013 28th IEEE/ACM
  International Conference on Automated Software Engineering (ASE)}.\hskip 1em
  plus 0.5em minus 0.4em\relax IEEE, 2013, pp. 367--377.

\bibitem{zhong2013exposing}
H.~Zhong, S.~Thummalapenta, and T.~Xie, ``Exposing behavioral differences in
  cross-language api mapping relations,'' in \emph{International Conference on
  Fundamental Approaches to Software Engineering}.\hskip 1em plus 0.5em minus
  0.4em\relax Springer, 2013, pp. 130--145.

\bibitem{lin2014detecting}
Y.~Lin, Z.~Xing, Y.~Xue, Y.~Liu, X.~Peng, J.~Sun, and W.~Zhao, ``Detecting
  differences across multiple instances of code clones,'' in \emph{Proceedings
  of the 36th International Conference on Software Engineering}, 2014, pp.
  164--174.

\bibitem{zhang2016isomorphic}
J.~Zhang, Y.~Lou, L.~Zhang, D.~Hao, L.~Zhang, and H.~Mei, ``Isomorphic
  regression testing: Executing uncovered branches without test augmentation,''
  in \emph{Proceedings of the 2016 24th ACM SIGSOFT International Symposium on
  Foundations of Software Engineering}.\hskip 1em plus 0.5em minus 0.4em\relax
  ACM, 2016, pp. 883--894.

\bibitem{xie2007towards}
T.~Xie, K.~Taneja, S.~Kale, and D.~Marinov, ``Towards a framework for
  differential unit testing of object-oriented programs,'' in \emph{Second
  International Workshop on Automation of Software Test (AST'07)}.\hskip 1em
  plus 0.5em minus 0.4em\relax IEEE, 2007, pp. 5--5.

\bibitem{petsios2017nezha}
T.~Petsios, A.~Tang, S.~Stolfo, A.~D. Keromytis, and S.~Jana, ``Nezha:
  Efficient domain-independent differential testing,'' in \emph{2017 IEEE
  Symposium on Security and Privacy (SP)}.\hskip 1em plus 0.5em minus
  0.4em\relax IEEE, 2017, pp. 615--632.

\bibitem{artzi2011framework}
S.~Artzi, J.~Dolby, S.~H. Jensen, A.~M{\o}ller, and F.~Tip, ``A framework for
  automated testing of javascript web applications,'' in \emph{Proceedings of
  the 33rd International Conference on Software Engineering}, 2011, pp.
  571--580.

\bibitem{pacheco2007feedback}
C.~Pacheco, S.~K. Lahiri, M.~D. Ernst, and T.~Ball, ``Feedback-directed random
  test generation,'' in \emph{29th International Conference on Software
  Engineering (ICSE'07)}.\hskip 1em plus 0.5em minus 0.4em\relax IEEE, 2007,
  pp. 75--84.

\bibitem{pradel2012fully}
M.~Pradel and T.~R. Gross, ``Fully automatic and precise detection of thread
  safety violations,'' in \emph{Proceedings of the 33rd ACM SIGPLAN conference
  on Programming Language Design and Implementation}, 2012, pp. 521--530.

\bibitem{chiang2014efficient}
W.-F. Chiang, G.~Gopalakrishnan, Z.~Rakamaric, and A.~Solovyev, ``Efficient
  search for inputs causing high floating-point errors,'' in \emph{Proceedings
  of the 19th ACM SIGPLAN symposium on Principles and practice of parallel
  programming}, 2014, pp. 43--52.

\bibitem{fraser2011evosuite}
G.~Fraser and A.~Arcuri, ``Evosuite: automatic test suite generation for
  object-oriented software,'' in \emph{Proceedings of the 19th ACM SIGSOFT
  symposium and the 13th European conference on Foundations of software
  engineering}.\hskip 1em plus 0.5em minus 0.4em\relax ACM, 2011, pp. 416--419.

\bibitem{alshahwan2011automated}
N.~Alshahwan and M.~Harman, ``Automated web application testing using search
  based software engineering,'' in \emph{Proceedings of the 2011 26th IEEE/ACM
  International Conference on Automated Software Engineering}.\hskip 1em plus
  0.5em minus 0.4em\relax IEEE Computer Society, 2011, pp. 3--12.

\bibitem{rojas2016seeding}
J.~M. Rojas, G.~Fraser, and A.~Arcuri, ``Seeding strategies in search-based
  unit test generation,'' \emph{Software Testing, Verification and
  Reliability}, vol.~26, no.~5, pp. 366--401, 2016.

\bibitem{fraser2012seed}
G.~Fraser and A.~Arcuri, ``The seed is strong: Seeding strategies in
  search-based software testing,'' in \emph{2012 IEEE Fifth International
  Conference on Software Testing, Verification and Validation}.\hskip 1em plus
  0.5em minus 0.4em\relax IEEE, 2012, pp. 121--130.

\bibitem{mcminn2012search}
P.~McMinn, M.~Shahbaz, and M.~Stevenson, ``Search-based test input generation
  for string data types using the results of web queries,'' in \emph{2012 IEEE
  Fifth International Conference on Software Testing, Verification and
  Validation}.\hskip 1em plus 0.5em minus 0.4em\relax IEEE, 2012, pp. 141--150.

\bibitem{alshraideh2006search}
M.~Alshraideh and L.~Bottaci, ``Search-based software test data generation for
  string data using program-specific search operators,'' \emph{Software
  Testing, Verification and Reliability}, vol.~16, no.~3, pp. 175--203, 2006.

\bibitem{godefroid2005dart}
P.~Godefroid, N.~Klarlund, and K.~Sen, ``Dart: directed automated random
  testing,'' in \emph{Proceedings of the 2005 ACM SIGPLAN conference on
  Programming language design and implementation}, 2005, pp. 213--223.

\bibitem{sen2005cute}
K.~Sen, D.~Marinov, and G.~Agha, ``Cute: a concolic unit testing engine for
  c,'' \emph{ACM SIGSOFT Software Engineering Notes}, vol.~30, no.~5, pp.
  263--272, 2005.

\bibitem{thummalapenta2011synthesizing}
S.~Thummalapenta, T.~Xie, N.~Tillmann, J.~De~Halleux, and Z.~Su, ``Synthesizing
  method sequences for high-coverage testing,'' \emph{ACM SIGPLAN Notices},
  vol.~46, no.~10, pp. 189--206, 2011.

\bibitem{xu2013mining}
W.~Xu, T.~Ding, H.~Wang, and D.~Xu, ``Mining test oracles for test inputs
  generated from java bytecode,'' in \emph{2013 IEEE 37th Annual Computer
  Software and Applications Conference}.\hskip 1em plus 0.5em minus 0.4em\relax
  IEEE, 2013, pp. 27--32.

\bibitem{dallmeier2010generating}
V.~Dallmeier, N.~Knopp, C.~Mallon, S.~Hack, and A.~Zeller, ``Generating test
  cases for specification mining,'' in \emph{Proceedings of the 19th
  international symposium on Software testing and analysis}, 2010, pp. 85--96.

\bibitem{della2017saying}
L.~Della~Toffola, C.-A. Staicu, and M.~Pradel, ``Saying ‘hi!’is not enough:
  Mining inputs for effective test generation,'' in \emph{2017 32nd IEEE/ACM
  International Conference on Automated Software Engineering (ASE)}.\hskip 1em
  plus 0.5em minus 0.4em\relax IEEE, pp. 44--49.

\end{thebibliography}
\bibliographystyle{IEEEtran}
%\begin{thebibliography}{1}
%\bibliography{bibliography}

%\end{thebibliography}

\begin{comment}

\begin{IEEEbiography}{Michael Shell}
Biography text here.
\end{IEEEbiography}

\begin{IEEEbiographynophoto}{John Doe}
Biography text here.
\end{IEEEbiographynophoto}

%\newpage

\begin{IEEEbiographynophoto}{Jane Doe}
Biography text here.
\end{IEEEbiographynophoto}


%\vfill

\end{comment}

\end{document}